\documentclass[apj]{emulateapj}

\newcommand {\SiII}   {\ion{Si}{2}}  
\newcommand {\SiIII}  {\ion{Si}{3}}  
\newcommand {\SiIV}   {\ion{Si}{4}}  
\newcommand {\HI}     {\ion{H}{1}}   
\newcommand {\HII}    {\ion{H}{2}}   
\newcommand {\OVI}    {\ion{O}{6}}   
\newcommand {\OI}     {\ion{O}{1}}   
\newcommand {\CIII}   {\ion{C}{3}}   
\newcommand {\CIV}    {\ion{C}{4}}   
\newcommand {\NV}     {\ion{N}{5}}    

\newcommand{\NHI}{$N_{\rm HI}$}
\newcommand{\HST}{{\it HST}}
\newcommand{\FUSE}{{\it FUSE}}
\newcommand{\kms}{km~s$^{-1}$}
\newcommand\etal{et~al.}
\newcommand{\cd}{cm$^{-2}$}

\begin{document}

\title{Hubble Space Telescope Survey of Interstellar High-Velocity \ion{Si}{3}}    

\author{Joseph A. Collins\altaffilmark{1}, J. Michael Shull}
\affil{University of Colorado, CASA, Department of Astrophysical 
\& Planetary Sciences, Campus Box 389, Boulder, CO 80309 \\ 
jcollins@casa.colorado.edu, michael.shull@colorado.edu}

\altaffiltext{1}{Also at Front Range Community College, Larimer Campus, 
4616 S. Shields St., Fort Collins, CO 80526}

\and

\author{Mark L. Giroux}
\affil{East Tennessee State University, Department of Physics \& Astronomy,
   \\ Box 70652, Johnson City, TN 37614, girouxm@etsu.edu}

\begin{abstract}

We describe an ultraviolet spectroscopic survey of interstellar 
high-velocity cloud (HVC) absorption in the strong $\lambda1206.500$ 
line of \SiIII\ using the Space Telescope Imaging Spectrograph aboard the 
{\it Hubble Space Telescope}.  Because the Si~III line is 4--5
times stronger than \OVI\ $\lambda1031.926$, it provides a sensitive 
probe of ionized gas down to column densities 
${\rm N}_{\rm SiIII} \approx 5 \times 10^{11}$ \cd\ at \SiIII\ equivalent 
width 10 m\AA. We detect high-velocity \SiIII\ over $91\pm4$\% of the sky 
(53 of 58 sight lines); 59\% of the HVCs show negative LSR velocities.  
Per sight line, the mean HVC column density is
$\langle \log{\rm N}_{\rm SiIII} \rangle = 13.19 \pm 0.45$, 
while the mean for all 90 velocity components is $12.92 \pm 0.46$.
Lower limits due to \SiIII\ line saturation are included in this average, 
so the actual mean/median values are even higher. 
The \SiIII\ appears to trace an extensive ionized component of Galactic 
halo gas at temperatures $10^{4.0-4.5}$~K indicative of a cooling flow.  
Photoionization models suggest that typical \SiIII\ absorbers with 
$12.5 < \log {\rm N}_{\rm SiIII} < 13.5$ have total hydrogen column 
densities N$_{\rm H} \approx 10^{18}$--$10^{19}$~\cd\ for gas of
hydrogen density $n_H \approx 0.1$ cm$^{-3}$ and 10\% solar metallicity.  
With typical neutral fractions 
${\rm N}_{\rm HI}/{\rm N}_{\rm H} \approx 0.01$, these HVCs may elude 
even long-duration 21-cm observations at Arecibo, the EVLA, and other
radio facilities.  However, if \SiIII\ is associated with higher density 
gas, $n_H \geq 1$ cm$^{-3}$, the corresponding neutral hydrogen could be 
visible in deep observations.  This reservoir of ionized gas may contain 
$10^8~M_{\odot}$ and produce a mass infall rate of 
$1~M_{\odot}~{\rm yr}^{-1}$ to the Galactic disk.     

\end{abstract}

\keywords{Galaxy: halo --- ISM: clouds --- ISM: abundances --- 
ultraviolet: general }

\noindent

\section{Introduction}

Over the past several decades, radio maps and spectra of \HI\ 
21-cm emission have revealed a significant population of objects with
velocities incompatible with standard models of Galactic rotation, the
so-called high-velocity clouds (Wakker \& van Woerden 1997). 
In the Milky Way, high-velocity clouds (HVCs) and intermediate-velocity 
clouds (IVCs) are distinguished by their velocities in the local
standard of rest (LSR), with a nominal population break at 
$|V_{\rm LSR}| = 90-100$ \kms.  Following the convention of 
Wakker \& van Woerden (1997), we define IVCs to lie between 
$|v_{\rm LSR}| \approx$ 30--90\,\kms\ and HVCs to have 
$|v_{\rm LSR}| > 90$~\kms.  

Far more sensitive probes of HVCs and IVCs have become available
through ultraviolet (UV) absorption-line studies of zero-redshift 
absorption lines in the spectra of background quasars taken with the 
{\it Hubble Space Telescope} (\HST) and the 
{\it Far Ultraviolet Spectroscopic Explorer} (\FUSE).  
The far ultraviolet offers much greater sensitivity to low column 
density gas, such as HVC surveys in \OVI\ (Sembach \etal\ 2003) and 
\SiIII\ (Shull \etal\ 2009), and they offer the best means for tracing 
ionized HVC gas.  The fractional HVC sky coverage using these lines 
range from 60 to 90\%, reflecting both the higher UV sensitivity and 
the fact that HVCs contain substantial amounts of ionized gas 
undetectable in 21-cm emission.  Consequently, the total reservoir of 
HVC gas is considerably larger than seen in \HI, and 
the cooling mass-infall rate could be as high as 
$1M_{\odot}$ yr$^{-1}$ (Shull \etal\ 2009).  This infall provides  
a significant portion of the replenishment rate required to fuel 
star formation in the Galactic disk.  On the scale of the Galactic halo, 
the G-dwarf problem (Pagel 1994) has inspired a picture of continuous 
infall of low-metallicity gas onto the Milky Way disk.

On larger scales, HVCs have been invoked to explain phenomena such as 
the ``missing dwarfs'' predicted in simulations of Local Group formation 
(Blitz \etal\ 1999) or the potential reservoir of hot baryons located 
in the circumgalactic medium of large galaxies (Nicastro \etal\ 2003) 
or in the true intergalactic medium (IGM). Whatever their distance, 
HVCs may trace building blocks of the Galaxy or Local Group 
(Putman \etal\ 2009).  Their study illuminates 
the processes of local gaseous infall/outflow, structure formation, and 
chemical enrichment of the Milky Way and IGM.
Most of the framework used to interpret emission and absorption from 
high-velocity gas comes from a thorough understanding of the kinematics, 
distribution, and composition of the \HI\ column density distribution.
At the high end of this distribution are those HVCs mapped through all-sky 
\HI\ 21-cm emission surveys such as the Leiden-Dwingleoo Survey 
(LDS; Hartmann \& Burton 1997), sensitive to column densities down to 
$N_{\rm HI} \approx 10^{18.5}$ to $10^{19}~$\cd\ for a 
standard HVC width of $\Delta V\sim100$ km s$^{-1}$.  The high-velocity 
\HI\ emission sky is dominated by objects of large angular size 
known as HVC Complexes.  Inferring a scaled mass as a function of 
distance (e.g., Wakker \etal\ 1999) restricts the large complexes 
to the Galactic halo.  Uncertain distances to HVCs made it difficult to 
distinguish between various models and the role of HVCs in Galaxy evolution.  
However, recent work searching for \ion{Ca}{2}\ absorption in spectra of 
Galactic halo stars puts the distance to HVC Complex C at $10 \pm 2.5$ kpc 
(Thom \etal\ 2008; Wakker \etal\ 2007).  Distance to other HVCs 
(Wakker \etal\ 2008) have been constrained to 5--15 kpc by this technique.  

Our group and others have devoted considerable attention to Complex C, 
which is pierced by 11 QSO sight lines observable with UV spectrographs
on \HST\ and \FUSE.  Based on \OI\ and \ion{S}{2}\ abundances, its 
metallicity has been measured at $Z/Z_{\odot} = 0.09-0.29$ 
(Wakker \etal\ 1999; Gibson \etal\ 2001; Richter \etal\ 2001; 
Tripp \etal\ 2003; Collins, Shull, \& Giroux 2003, 2007, hereafter CSG03 
and CSG07).  These surveys found that Complex~C contains 
$\sim10^7~M_{\odot}$ of gas, with approximately equal amounts of
neutral and ionized components.   Some of this HVC is mixing with 
Galactic gas as it falls toward the disk.  At an approximate
elevation of 10~kpc and infall velocity of $-150$~\kms, the gas in
Complex~C will fall onto the Galactic disk over the next 50--100 Myr,
with an average infall rate of $0.1~M_{\odot}~{\rm yr}^{-1}$.   
Fox \etal\ (2004) and CSG07 investigated the highly ionized portions 
of Complex~C, reinforcing the notion that the cloud is interacting with 
the Galactic halo through cooling and dynamic instabilities.

While ever-deeper \HI\ emission studies have pushed the catalog 
of known HVCs to smaller ${\rm N}_{\rm HI}$, the most dramatic strides 
in our knowledge of the low end of the HVC column density distribution 
have come from UV spectroscopy. The \FUSE\ survey (Sembach \etal\ 
2003, hereafter S03) of \OVI\ detected high-velocity absorption
in $60-85$\% of extragalactic sight lines. Our \HST\ study in the strong
line of \SiIII\ (Shull \etal\ 2009) found HVCs in 30 of 37 sight lines
(approximately 80\%).  Although many of these high-velocity \OVI\ and 
\SiIII\ detections correlate directly with \HI-detected HVCs, a 
significant portion of the population has no corresponding 21-cm 
detections.  The \OVI\ HVCs not detected in \HI\ emission have been 
termed ``highly ionized HVCs'' (Sembach \etal\ 1999).  However, 
our recent studies of this population (Collins, Shull, \& Giroux 2004, 
2005; hereafter CSG04, CSG05) have detected neutral and low-ionized gas
at the same velocities.  We suggest that this HVC population is simply an 
extension of the usual HVC distribution, but with \HI\ column densities 
below the sensitivity level of current 21-cm emission surveys. 

One interpretation (Nicastro \etal\ 2002; 2003) of the kinematics and 
ionization properties of the highly ionized HVCs associates these 
absorbers with warm-hot \ion{O}{7}\ and \ion{O}{8}\ absorbing IGM 
filaments or WHIM (Cen \& Ostriker 1999) in the Local Group.
We have argued (CSG04, CSG05) that these HVCs more likely reside in 
the Galactic halo, based on three major lines of evidence.  First, the 
mass in hot gas would be unreasonably large, 
$M_{\rm hot}\sim[10^{12} M_{\odot}]d_{\rm Mpc}^2$ (see CSG04), 
compared to dynamical measures of the Milky Way and Local Group 
if all the \OVI\ absorbers traced a WHIM filament at Mpc distances in 
the Local Group.  Distributing the HVCs throughout the Local Group only 
lowers this estimate by factors of 2--3.  
Second, the kinematics of HVCs are consistent with a combination 
of infall onto the Galaxy and outflows in other directions
In the current sample, as well as that of Shull \etal\ (2009),
about 60\% of the HVCs have negative LSR velocities. 
Third, 10 of the 12 highly-ionized HVCs surveyed in CSG05 contain 
neutral gas or low ions (\ion{C}{2}, \ion{Si}{2}, \ion{Si}{3}),
suggesting that these HVCs have densities $n_H \gg 10^{-4}$ cm$^{-3}$   
to avoid being photoionized by metagalactic radiation.  We argue
therefore that these HVCs are similar in ionization properties to
the population observed in 21-cm emission, but with lower neutral
hydrogen column densities, log~\NHI\ $\leq 18$.  The higher ionization
states (\OVI, \CIV, \SiIII, etc.) seen in these HVCs suggests that
they contain multiphase gas, produced by both photoionization and
collisional ionization (shocks or interfaces with hotter halo gas).   
A recent survey (Fox, Savage, \& Wakker 2006; hereafter FSW06) found 
that high-velocity \ion{C}{3}\ is present in $\sim80$\% of the 
highly-ionized HVCs, consistent with a multiphase interpretation 
of these HVCs and a Galactic origin.  This interpretation is also
supported by Zech \etal\ (2008) who analyzed a highly ionized HVC 
toward the Globular Cluster M5 (7.5 kpc distance).  Given the 
similarities in ionization properties between the highly ionized
HVCs and the 21 cm-detected halo HVCs, a more apt term for the 
highly ionized HVCs may be ``low column HVCs''.

The full distribution, kinematics, and covering fraction of the low column 
density HVCs is not well determined.  Although \ion{H}{1} Ly$\alpha$ 
and higher Lyman-series resonance
lines are the most sensitive diagnostics of low column density gas, the 
absorption in high-velocity gas is rarely open to analysis in these lines,
owing to severe saturation from Galactic \ion{H}{1} absorption and confusion 
with other resonance lines.  Considering elemental abundance, ionization, and 
oscillator strength, the \ion{Si}{3} $\lambda1206.50$ line 
is one of the best (strongest) UV resonance lines for discovering 
high-velocity gas.  Although the 
\ion{Si}{2} $\lambda1260.42$ and \ion{C}{2} $\lambda1334.53$ lines are 
typically just as strong, the \ion{S}{2} $\lambda1259.52$ and 
\ion{C}{2}* $\lambda1335.70$ lines contaminate large portions of their 
high-velocity profiles. An additional advantage of the \ion{Si}{3} line 
is that it lies in a relatively line-free part of the spectrum, blueward
of Ly$\alpha$, uncontaminated by the intergalactic Ly$\alpha$ forest, and 
in most cases, by other IGM lines.  

A large number of extragalactic AGN/QSO sight lines have been observed
with the Space Telescope Imaging Spectrograph (STIS) on \HST.
These STIS archive spectra cover the $\lambda=1200$ \AA\ region, enabling
us to use the \ion{Si}{3} line to map the distribution, covering fraction, 
and kinematics of HVCs down to total hydrogen column densities 
${\rm N}_{\rm H} \approx 10^{17}$~\cd, and below the detection limits 
typically available in \HI.  In this paper, we present a survey of 
high-velocity interstellar \SiIII\ absorption 
line along 58 extragalactic sight lines, using archival data taken with 
the \HST-STIS E140M echelle and G140M gratings.  A previous paper
(Shull \etal\ 2009) presented a study of \SiII, \SiIII\, and \SiIV\ 
in HVCs and IVCs in the E140M datasets. The survey data and criteria for 
their selection are presented in Section~2.  The analysis of those data, 
along with line profiles and measurements, is discussed in Section~3.  
We explore the overall distribution, kinematics, and ionization of these 
HVCs in Section~4.  Our conclusions are presented in Section 5.

\section{Sight Line Selection and Survey Data}

The initial pool of data for this study is the entire archive of 
\HST-STIS spectra of extragalactic targets that cover the \SiIII\ 
line.  Since we aim to measure absorption features of width 
$\sim100$ km~s$^{-1}$, we only use data taken with the E140M echelle or 
G140M first-order gratings.  The E140M mode provides 7 \kms\ resolution 
over a large bandpass,
1150 -- 1700 \AA, which covers resonance lines of several important ion
species.  The G140M grating, when set to cover the 1199--1249 \AA\ range,  
provides higher sensitivity at the cost of smaller bandpass and lower 
resolution ($\sim25$ km s$^{-1}$) compared to the echelle data.  
In order to improve detectability of low column density absorbers and to match
the resolution provided by the G140M data, we binned all E140M data sets to
three pixels.  Some sight lines
contain both E140M and G140M data covering the \ion{Si}{3} line.  In these 
cases, we rely on the E140M datasets, since they provide the opportunity 
to compare a high-velocity feature in multiple absorption lines and facilitate
the identification of IGM and other contaminants.

Once we compiled this set of spectra, we narrowed our survey with a few 
criteria.  First, we limited our survey to datasets with 
sufficient signal to noise to analyze high-velocity features: 
$(S/N)\geq3$ at $1200$ \AA\ per resolution element (three binned pixels) 
for the E140M data and per pixel for the G140M data.  Although lower 
quality data may identify high-velocity features, the continuum is often
too noisy to properly establish its velocity extent.  Second, we eliminated  
sight lines from the survey in which absorbers at redshifts $z>0$ 
significantly contaminate the high-velocity range of the \SiIII\ line.  
Two sight lines dropped from our survey, 3C~249.1 and Mrk 205, 
deserve mention.  High-velocity \ion{Si}{3} at negative velocity is most likely
present in each sight line, as evidenced by corresponding detections in other 
absorption lines.  However, the HVC feature is contaminated by intergalactic
Ly$\beta$ (at $z = 0.176$) toward 3C~249.1 and by intergalactic \ion{N}{1} 
$\lambda1200.71$ (at $z = 0.0043$) from a strong Ly$\alpha$ absorber toward 
Mrk 205.  

After we applied the selection criteria, our survey consisted of 58 sight 
lines, of which 33 and 25 are E140M and G140M datasets, respectively.  
Absolute wavelength scales were obtained by aligning the peak of 
\ion{H}{1} emission features in sight line data from the Effelsberg 100-m 
telescope (Wakker \etal\ 2003) or the Leiden-Dwingleoo Survey (LDS; 
Hartmann \& Burton 1997), when available,
with corresponding absorption line features in the STIS data.  We generally 
relied on lines of \ion{N}{1} ($\lambda$1199.55, 1200.22, 1200.71) and 
\ion{S}{2}\ ($\lambda$1250.58, 1253.81, 1259.52) for this comparison.
In a few cases, we could not find an absolute wavelength scale, owing to 
absence of \ion{H}{1} data, or in a few G140M datasets when the available 
comparison features were not clean enough to use.

\section{Measurement of High-Velocity \ion{Si}{3} Features}

We now discuss our techniques for investigating \SiIII\ absorption from 
HVCs in our sample of 58 sight lines.  In order to measure column densities 
of ion species, we first extracted individual line profiles from the full 
\HST-STIS spectra.  These profiles were normalized by fitting low-order 
polynomials to the continuum $\pm$(3--10)~\AA\ about the line in question, 
although in a number of cases spurious features near the line required using 
a much larger region for continuum measurement.  For each sight line, we
searched the region $100 < |v| < 500$ km s$^{-1}$ around the \SiIII\ line 
for any possible absorption from an HVC.  If candidate HVC features were 
found, we attempted to determine whether the feature was due to a
contaminant line at redshift $0<z<z_{QSO}$.  This process was more
easily performed for E140M spectra, where redshifts of any intervening
Ly$\alpha$ absorbers can be determined along with possible
contamination of the \ion{Si}{3} profile from other absorption lines
at the same redshifts.  Additionally, several of these sight lines have been 
examined in the literature for intervening lines, which can also 
be used to assess 
possible contamination of the \ion{Si}{3} line.  If contamination from
known IGM absorbers can be ruled out, we can then assume that any 
detected absorption is due to \ion{Si}{3}. 

Once we have identified high-velocity \SiIII\ absorbers, we determined 
the relevant velocity range by comparing the \SiIII\ profile to other 
absorption lines in which the high velocity component is detected.  
In some cases, the \HI\ emission profile can be used to identify the 
velocity range.  The most useful lines for the comparison, however, are
the strong lines of \ion{Si}{2} $\lambda1260.42$ and \ion{C}{2}
$\lambda1334.53$, although in many cases the HVCs can also be detected
in \ion{O}{1} $\lambda1302.17$, the \ion{N}{1} triplet at 1199--1200~\AA, 
other \ion{Si}{2} lines (e.g., 1193.29~\AA, 1304.37~\AA, 1526.71~\AA), 
and in the more highly ionized doublets of
\ion{C}{4} $\lambda\lambda1548.20, 1550.77$ and \ion{Si}{4}
$\lambda\lambda1393.76, 1402.77$.  Some of the sight lines are shared
by the \FUSE\ survey of high-velocity \ion{O}{6} (S03), with which 
high-velocity \SiIII\ features can be compared.  The presence of a 
high-velocity feature in multiple absorption lines provides 
additional confidence that the \SiIII\ profile is uncontaminated by
higher redshift absorption features.  Some G140M datasets cover only 
1199--1249 \AA, limiting our ability to establish an accurate integration 
range, confirm detections, and search for higher-redshift contaminants.  

Figures 1 and 2 show \SiIII\ profiles from E140M and G140M data,
respectively, for all 58 sight lines surveyed.  Also shown are
the velocity extents of each high-velocity component analyzed.
Once an integration range is established, we measure the equivalent width, 
$W_{\lambda}$, of the high-velocity feature.  
Using the most recent value of oscillator strength for the \SiIII\ line
($f=1.63$; Morton 2003), we measured the column density of high-velocity 
\SiIII\ using the apparent optical depth method (AOD; Savage \& Sembach 1991),
which applies to lines that are optically thin.  The linear relation between
Si~III equivalent width and column density is
$W_{\lambda} = (21.0~{\rm m\AA})({\rm N}_{\rm SiIII} / 10^{12}~{\rm cm}^{-2})$. 
Since the \ion{Si}{3} line is so strong, saturation, both resolved and 
unresolved, is often an issue.  
For a Gaussian profile, the \SiIII\ optical depth at line center is
$\tau_0 \approx (0.295) N_{12} b_{10}^{-1}$, scaled to a column density
${\rm N}_{\rm SiIII} = (10^{12}~{\rm cm}^{-2}) N_{12}$ and to a 
doppler width
$b_{\rm SiIII}$ = (10~\kms)$b_{10}$, chosen to reflect the turbulence and
velocity components that often broaden HVC metal-line absorbers. 
Typical HVCs exhibit \SiIII\ absorption over an extended velocity range,
$\Delta v = 40-100$ \kms, far broader than any single component would allow.
Indeed, the curves of growth derived for low metal ions in HVC Complex C
(CSG03, CSG07) range from $b \approx$ 7--18 \kms, consistent
with non-thermal line widths.  Given these large velocity dispersions, we
expect \SiIII\ line saturation to set in at column densities
N$_{\rm SiIII} > (3.4 \times 10^{12}~{\rm cm}^{-2})b_{10}$ in each
STIS/E140M resolution element.  

In cases where saturation is clearly present in a line profile, we 
adopt the measured AOD value as a lower limit to N(\ion{Si}{3}).  
The E140M data have a resolution of 7 km s$^{-1}$, allowing us to
search for saturation in the unbinned data.  If saturation is present 
in only one pixel in unbinned E140M data, its effects will not
be apparent in the data binned to three pixels.  In these cases, we adopt the 
AOD value as a lower limit.
Searching for unresolved saturation in the final G140M profiles is
more difficult, because those data are analyzed at their full resolution.  
To estimate the likelihood of unresolved saturation in HVCs detected in
G140M data we begin by examining the normalized flux at the trough of all 
high-velocity \SiIII\ features in the
binned E140M datasets.  We then determine whether saturation is present
in those features in the unbinned data.  For normalized fluxes 25\% or 
below at the absorption-line trough in the binned data, 
saturation is present in the unbinned data in almost all cases.  
For fluxes above that level, saturation is absent in all but a couple of cases.
The G140M data have a similar
resolution to the E140M data binned to three pixels, so we adopt 0.25
as the trough normalized flux, below which saturation is likely.  
In those cases where an HVC has a normalized flux below 0.25 at its trough,
we adopt the AOD value as the lower limit to N$_{\rm SiIII}$. 
This approach gives reasonable confidence in detecting saturation down to
resolutions of 7~\kms, but reveals little information on saturation 
over more narrow velocity ranges.

Tables 1 and 2 show the \SiIII\ column density measurements of all 90 
high velocity components detected in 53 of 58 surveyed sight lines, 
including target Galactic coordinates, velocity range of the HVC, and 
equivalent width for the E140M and G140M datasets, respectively.  In 
only a few sight lines did we fail to detect high-velocity \ion{Si}{3}.  
In order to determine a ($3 \sigma$) upper limit to high-velocity 
\SiIII\ in those sight lines, we first determined whether 
high-velocity gas was present in any other absorption lines.  
In two sight lines, 3C 273 and PKS 1302-102, high-velocity \OVI\ has 
been detected (S03; CSG05) and we adopted the \OVI\ velocity range 
to establish an upper limit on \ion{Si}{3} features.  If no HVC of
any kind was detected in a sight line, we assumed a typical HVC 
velocity width of 100 km s$^{-1}$ to establish upper limits. 
Typically, there is little difference in signal-to-noise (S/N) ratio 
across 4--5 \AA, except for the Mrk~1383 profile where the blueward
side is slightly more noisy.  
Table 3 shows these 5 sight lines where high-velocity \SiIII\ is
not detected, along with upper limits on ${\rm N}_{\rm SiIII}$.
Notes on individual sight lines from this survey are presented
in Appendices A and B.

\section{DISCUSSION}

\subsection{Distribution and Covering Factor}

From our sample of 58 sight lines, we can make an estimate of the sky covering
factor of high-velocity \ion{Si}{3}\ absorption.  We detected high-velocity
\ion{Si}{3} in 53 of the 58 sight lines, translating to a covering factor
of $91\pm4$\%.  In total, we found 90 HVC velocity components. 
The error bar corresponds to the possibility that we
have misinterpreted Si~III absorption in $\pm2$ sight lines.     
Interestingly, high-velocity \ion{Si}{3} is detected in all 25
of the sight lines with G140M data.  In the targets covered by E140M data, 
we detected high-velocity \ion{Si}{3} in 28 of 33 sight lines or $85\pm6$\%.  
Since there are several low-significance detections of weak HVCs, 
a better estimate of sky covering factor can be obtained by choosing a column
density cutoff.  If we choose a cutoff of $\log {\rm N}_{\rm SiIII} = 12.50$ for the 
combined HVC column density per sight line, the covering factor drops
to $84\pm4$\% (49 of 58), with a detection rate of $82\pm6$\% (27 of 33) 
in the E140M datasets and $88\pm8$\% (22 of 25) in the G140M datasets.
These \SiIII\ covering factors agree with the value of $81 \pm 5$\% 
found in our previous paper (Shull \etal\ 2009).  
The HVC sky covering factor has been investigated in other ions as well.  
The \FUSE\ survey (S03) of high-velocity \OVI\ found a covering factor of 
59\% in the full survey, and 85\% if one considers only the highest-S/N
datasets.  Where higher column density gas was probed, the covering
factor was measured at 37\% down to column density 
N$_{\rm HI} \gtrsim7\times10^{17}$ \cd\ (Murphy, Lockman, \& Savage 1995).
This lower covering \HI\ fraction suggests that the HVCs have extended
atmospheres of ionized gas, which is best probed by UV resonance lines
such as \SiIII\ or \OVI.



Although the high HVC detection rate of \SiIII\ indicates widespread
coverage of ionized gas on the sky, there is some evidence that the gas can be
clumpy on small scales.  Consider the sight-line pair 3C 273 and Q 1230+0115, 
separated by less than 1$\arcdeg$.  Towards Q1230+0115, there are two narrow
HVC features that can be detected in \ion{Si}{3}, as well as in other absorption 
lines.  These features are absent in the 3C 273 spectrum, which is among the 
highest-S/N spectra in this survey.  The narrowness of these absorption
features and their absence in a spectrum along a nearby sight line suggests
that they are relatively small clumps.  Located 5$\arcdeg$ away from this close 
pair is the sight line to PG 1216+069, which shows positive high-velocity 
absorption in multiple absorption lines, but with a completely different 
velocity profile than the HVCs toward Q1230+0115.

In several other cases, close sight line pairs show similar absorption, 
indicating that in many cases the high-velocity gas is more extended.  
The sight lines to Mrk 1044 and NGC 985 are separated by 1.1$\arcdeg$.  
Each sight line shows weak negative high-velocity \ion{Si}{3} over 
$-250<V_{\rm LSR}<-150$ \kms, with equivalent widths of
$79 \pm 17$ m\AA\ and $51 \pm 10$ m\AA, respectively.  The absorption 
toward NGC 985 having a slightly wider feature over two components.
The sight lines to PG 1049-005 and PKS 1103-006 are separated by
3.7$\arcdeg$, each showing very strong positive high-velocity features
with similar velocity structure in their \ion{Si}{3}\ profiles.

\subsection{The HVC Complexes}
One of the more extended high-velocity features in the sky is the
collection of HVCs in the quadrant of the sky comprising Complexes
A, C, K, and M, along with the Milky Way's Outer Arm.  
These Complexes are all roughly connected in velocity, extending out 
beyond $V_{\rm LSR}\approx-200$ \kms.  It is fortuitous that this part 
of the sky is pierced by numerous strong extragalactic sources for 
absorption-line studies.  Figure 3 plots the sight line locations in 
this survey for the region of sky $180\arcdeg\geq l \geq 50\arcdeg$ and
$70\arcdeg\geq b \geq 20\arcdeg$ together with contours of $N_{\rm HI}$ 
from the LDS over the velocity range $-210<V_{\rm LSR}<-90$ \kms.  
In the plotted region we have investigated 19 sight lines, nearly one-third
of the total from this survey.  Even though \ion{H}{1}\ 21-cm emission
is detected in only 9 of the 19 sight lines at 
${\rm N}_{\rm HI} \geq 1\times10^{19}$ \cd\ 
(roughly the sensitivity of the LDS) over the velocity range
$-210 < V_{\rm LSR} < -90$ \kms, we detected high-velocity \SiIII\ in
all 19 of these sight lines over some portion of that velocity range.  
Even the sight lines to PG 1444+407 and Mrk 478, which are separated 
by $\sim5\arcdeg$ and lie $\gtrsim15\arcdeg$ from any region of Complex~C 
with ${\rm N}_{\rm HI} > 10^{19}$~\cd, show clear \SiIII\ absorption at
similar velocity.  It should be noted, however, that NGC 4051, one of 
the sight lines without high-velocity \ion{Si}{3}\ absorption, lies just 
outside the map boundary at $(l,b)=(148.89^{\circ},70.09^{\circ})$ 
within a few degrees of 21-cm emission from Complex M.  

The extended \SiIII\ absorption probably traces thinner 
ionized portions of these extended HVC complexes, or a hotter envelope 
outside the cooler regions of high ${\rm N}_{\rm HI}$.  The second
possibility seems to be contrary to the distribution of \OVI\ observed 
by FSW06.  They claim significant non-detections of \OVI\ along sight 
lines only a few degrees away from the 21-cm emission of these HVC 
Complexes, suggesting that the hot layer is closely confined to 
the Complexes.  We confirm this trend along 19 sight lines near the 
HVC Complexes shown in Figure 3.  We examined 10 sight lines in this 
region that contain \SiIII\ HVCs at velocities similar to the HVC 
Complexes, yet outside the ${\rm N}_{\rm HI} = 1\times10^{19}$ \cd\ contour. 
Of five sight lines with viable \FUSE\ data covering \OVI\ (PG 1444+407, 
Mrk 478, H 1821+643, VII~Zw118, PG 0953+414), we detect high-velocity
\OVI\ at the same velocity in only one case (H 1821+632). 
In the other four sight lines containing high-velocity \SiIII\ but 
no \OVI, the \SiIII\ detections are weak.  If \SiIII\ is entrained 
in an extended hot envelope to the HVC Complexes, one possible
explanation for the lack of \OVI\ may simply be the lower sensitivity 
to high-velocity gas in the weaker \OVI\ line.  (The \FUSE\ data
generally have lower S/N than STIS.)   
Alternatively, if the hot \OVI-bearing gas is confined to the 
vicinity of the HVC Complexes, then the \SiIII\ gas may trace
much lower column density regions in the HVC Complexes.
 
The fact that 19 of the 58 (33\%) sight lines considered in this study lie
within a region covering only $\sim10\%$ of the sky does not significantly
bias our results.  Although we calculate a $91\pm4$\% covering factor based on 
the full sample of 58 sight lines, the covering factor only drops to 
87\% if we ignore the sight lines within the region plotted in Figure 3.

\subsection{Kinematics}

Another topic that can be explored with this survey is the kinematics
of the detected \ion{Si}{3}\ HVCs.  In Figure 4 we show the locations
of the 58 sight lines analyzed in this study, color-coded according to
whether the detected HVCs are blueshifted or redshifted.  This plot 
includes five sight lines where no high-velocity \ion{Si}{3}\ is detected.  
Also labeled are the Local Group barycenter at
$(\ell, b) = (147\arcdeg,-25\arcdeg)$ (Karachentsev \& Makarov 1996) and
the direction of the Milky Way's motion, $(\ell, b) = (107\arcdeg,-18\arcdeg)$
at $V=90$ \kms\ (Einasto \& Lynden-Bell 1982) with respect to the
Local Group barycenter.  

As seen in our previous studies of HVC kinematics (CSG05) and noted by 
Shull \etal\ (2009), the \ion{Si}{3}\ HVC kinematics shows a segregation 
by velocity in a dipole pattern according to Galactic longitude.  With a few 
exceptions, \ion{Si}{3} HVCs at negative velocity are at $0\arcdeg<l<180\arcdeg$
and \ion{Si}{3}\ HVCs at positive velocity are at $180\arcdeg<l<360\arcdeg$.  
A similar dipole pattern has been observed in high-velocity \ion{O}{6}\
(S03), highly-ionized HVCs (CSG05, FSW06), and in high-velocity \ion{H}{1}. 
Such a pattern can be explained to first order, by several models that decouple 
the HVCs from rotation of the Galactic disk.  CSG05 showed that a simple model 
of infall from a non-rotating spherical shell of clouds produces the general 
segregation trend, but not the absolute LSR velocities.
Models with no lag do not work.  If the clouds were rotating with the galaxy,
they would not be classified HVCs in the first place.  

Most previous HVC models assume that the circular velocity remains
constant on cylinders, as one moves above the Galactic plane.  A velocity 
lag has recently been noted in the Sloan Digital Sky Survey of $\sim$200,000
F and G stars (Ivezi\'c \etal\ 2008), and a circular velocity lag in halo
rotation has also been seen in external spiral galaxies (Collins,
Benjamin, \& Rand 2002).  The general pattern can be explained by
any of several models in which the HVCs are unconnected, or at
least nearly so, to disk rotation.  Simple models that produce this dipole
pattern include radial infall from non-rotating HVCs, or a slowing
rotation rate with height of the Galactic Disk. 

\subsection{Ionization Modeling}

Figure 5 shows the \SiIII\ column densities of all 90 HVC velocity 
components analyzed here vs.\ Galactic longitude and latitude.  
No clear trends are seen, other than possible peaks in 
\SiIII\ column densities around $\ell = 80-120^{\circ}$ and 
$\ell = 260-300^{\circ}$.  The line saturation in stronger
components makes it difficult to make more specific identifications
of these features.  Studies of weaker lines from other ions
should help.   

Averaging high-velocity \SiIII\ absorption over all sight lines, 
we found a mean column density 
$\langle \log{\rm N}_{\rm SiIII} \rangle = 13.19 \pm 0.45$ 
(median 13.29).  If we treat each velocity component independently, 
we find $\langle \log{\rm N}_{\rm SiIII} \rangle = 12.92 \pm 0.46$
(median 12.91).  Thus, the typical HVC seen in \SiIII\ has a column 
density $12.5 < \log{\rm N}_{\rm SiIII} < 13.5$.  We scale to a 
metallicity of 20\% solar, adopting a solar abundance 
(Asplund \etal\ 2005) of (Si/H)$_{\odot} = 3.24 \times 10^{-5}$.  
The median \SiIII\ column density corresponds to 
N$_{\rm HII} \approx (6 \times 10^{18}~{\rm cm}^{-2})
(Z_{\rm Si}/0.2Z_{\odot})^{-1}$ as the column of ionized 
low-metallicity gas.  These values are similiar to those inferred 
from the HVCs observed in \OVI\ (Sembach \etal\ 2003).

In a recent paper, Shull \etal\ (2009) analyzed a subset
of 83 HVC and IVC absorbers with \SiII, \SiIII, and \SiIV\ data
toward 37 active galactic nuclei at high latitude.  They detected
interstellar Si~III $\lambda1206.50$ absorption in numerous
high-velocity clouds (61 HVCs along 30 sight lines) and
intermediate-velocity clouds (22 IVCs along 20 sight lines).
The fractional HVC sky coverage is large ($81 \pm 5$\%
for 30 out of 37 directions) with \SiIII\ optical depth typically
4--5 times that of \OVI\ $\lambda 1031.926$.
By assuming that \SiII, \SiIII, \SiIV, and \HI\ are photoionized, 
Shull \etal\ (2009) constrained the mean photoionization parameter in 
the low halo to be $\langle \log\,U \rangle = -3.0^{+0.3}_{-0.4}$, 
approximately ten times lower than in the low-redshift IGM
(Danforth \& Shull 2008).  
The metallicities in a subset of these HVCs derived from [\SiII/\HI]
and [\OI/\HI] are 10--30\% solar, whereas values found from all 
three silicon ions are lower in the pure-photoionization models.  
These formally lower metallicities, 
$\langle \log\,(Z_{\rm Si}/Z_{\odot})\rangle = -2.1^{+1.1}_{-0.3}$
in 17 HVCs and $-1.0^{+0.6}_{-1.0}$ in 19 IVCs, are highly uncertain, 
since some of the higher ions may be collisionally ionized.  Thus, 
we see no reason to challenge the more reliable values from \OI\
and \SiII. 

The larger HVC covering factor seen in \SiIII\ absorption (80--90\%) 
compared to \HI\ emission (37\%) suggests that HVCs have extended 
atmospheres of ionized gas, probably of lower density and shredded by 
HVC motion through a hot halo medium.  Thermal interfaces between the 
\HI\ cores and the hotter substrate (Fox \etal\ 2006; CSG04; CSG05) 
may be responsible for the higher ions (\OVI, \CIV, \NV).  Thus, 
comparing probes of the ionized gas 
(\OVI, \SiIII) with 21-cm maps (\HI) could be quite helpful in understanding
the cloud geometry.  Significant advances in HVC studies could therefore
come from improvements in 21-cm sensitivity, reaching below the current
limits of ${\rm N}_{\rm HI} \approx 10^{18}$~\cd.  
As the \HI\ column density within an absorber falls below
$10^{18}$~cm$^{-2}$, the effects of self-shielding are reduced, and   
it is more straightforward to apply photoionization models to obtain 
better estimates of the physical properties of the gas.  In many 
plausible photoionization scenarios, the expected column density of 
neutral hydrogen accompanying \SiIII\ absorbers can exceed $10^{17}$~\cd. 

In Figure 6, we plot the results of two photoionization models
consistent with \SiII, \SiIII, and \SiIV\ column densities.  These   
models were constructed using the photoionization equilibrium code 
{\it Cloudy} (Ferland 1998).  Figure 6a shows an updated version of a 
model from Collins \etal\ (2003), who observed HVCs in Complex~C 
with ${\rm N}_{\rm HI} = 10^{19.0-20.1}$~\cd.  We modeled a 
constant-density slab of gas illuminated on one side by a hot-star 
spectrum, as might be expected for gas near the Galactic plane. 
The level of incident radiation is chosen so that the ionization 
parameter $\log U \approx -3.9$, and we assumed a metallicity of 
20\% solar.  The curves represent the cumulative calculated \SiII\ and
\SiIII\ column densities as one goes into the slab from the illuminated 
side, as measured by the cumulative \HI\ column density.  As expected, 
the cumulative Si~III column density asymptotes, as the Si~III ionizing 
photons ($h \nu \geq 16.34$ eV) are attenuated.  Once the slab becomes 
largely neutral in hydrogen, the dominant ionization species \SiII\ rises
linearly with cumulative column density.  As in previous photoionization 
models (Collins \etal\ 2003), we find neutral fractions 
${\rm N}_{\rm HI} /{\rm N}_{\rm H} \approx 0.4-0.7$  over the column-density 
range $19 < \log {\rm N}_{\rm HI} < 20$.  In this model, 
$\log {\rm N}_{\rm SiIII}$ reaches $13.1$ when the neutral hydrogen column 
${\rm N}_{\rm HI} \approx 10^{18}$ \cd.  The total hydrogen column is 
larger, N$_{\rm H} \approx 12$ \NHI. 

Figure 6b shows an ionization model consistent with those of Shull \etal\ 
(2009).  The level of incident radiation has an ionization parameter 
$\log U \approx -3.1$, and we assumed a metallicity of 10\% solar.  As in 
Figure 6a, the curves represent the cumulative column densities as one 
moves into the slab.  In this model, 
$\log {\rm N}_{\rm SiIII}$ reaches $13.5$ when 
$\log {\rm N}_{\rm HI} \approx 18$, and the total hydrogen column is
even larger than in Model (6a), with 
${\rm N}_{\rm H} \approx 70 {\rm N}_{\rm HI}$.
As these models indicate, the \SiIII\ HVC absorbers with 
$\log {\rm N}_{\rm SiIII} > 14$ likely require gas of lower 
density if the \SiIII\ is photoionized.  The weakest detectable
\SiIII\ absorbers with $\log {\rm N}_{\rm SiIII} \le 12$,
may be associated with higher density gas, or diffuse gas with 
${\rm N}_{\rm HI} < 10^{17}$ \cd.   In fact, regardless of density, the 
weak \SiIII\ absorbers are likely associated with \HI\ column densities
below the current detection limits for \HI\ 21-cm emission.

The $\log U = -3.1$ model (Figure 6b) is more consistent with the
analysis and ionization models of Shull \etal\ (2009).  In those\
studies, the \SiIII\ HVC absorbers with 
$\log {\rm N}_{\rm SiIII} = 12.5$ and 13.5 correspond to neutral
column densities ${\rm N}_{\rm HI} = 5 \times 10^{15}$~\cd\ and 
$6 \times 10^{16}$ \cd, respectively.  
The lower neutral fractions, ${\rm N}_{\rm HI}/{\rm N}_{\rm H}$, 
compared to the Complex-C models of CSG03 arise because the \HI\ column 
density in those models was much larger and the neutral fraction higher.  
The assumed gas densities in the CSG03 models were a factor of 4 higher.  
Although estimates of ionized gas scaled from \SiIII\ (Shull \etal\ 2009) 
remain reasonably similar for assumed values of metallicity and 
photoionization conditions, the corresponding \HI\ may vary considerably.  
In these constant-density {\it Cloudy} models, the ionizing flux is 
slowly attenuated with column density, but the neutral fraction 
remains rather constant at log~\NHI $\leq 17.5$.  Therefore, just as 
\SiIII\ is proportional to \HII, it also remains proportional to \HI\ 
at low column densities.

\section{CONCLUSIONS}

For this work, we have assembled all known \HST-STIS E140M and G140M 
datasets for extragalactic targets covering the \ion{Si}{3}$\lambda1206.50$
line.  This line is one of the strongest for HVC studies, and it lies in 
a part of the spectrum that is typically uncontaminated by higher-redshift 
IGM absorption systems.  After applying selection criteria, our sample was 
narrowed to 58 AGN targets.  Of those datasets, 33 were STIS/E140M spectra 
covering 1150--1700~\AA\ and 25 were G140M spectra that covered some portion 
of the spectrum around the \ion{Si}{3}\ line.  We measured velocity extents 
and \ion{Si}{3}\ column densities for a total of 90 detected HVC 
absorption features. From those measurements, we arrived at the 
following conclusions:
\begin{enumerate}

\item {\it Ultraviolet absorption spectra provide the most sensitive probes 
of ionized HVCs.}  We can detect high-velocity \SiIII\ in low column density 
(${\rm N}_{\rm HI} \approx 10^{17}$ \cd) portions of the HVC Complexes,
well below current senstivity levels in 21-cm emission.  
 
\item{\it The sky covering fraction of high-velocity \ion{Si}{3}\ 
$\lambda1206.5$ absorption is 80--90\%.}  This covering fraction
is considerably higher than that (37\%) seen in 21-cm emission.  
Because of its intrinsic strength, high-velocity \ion{Si}{3}\ is 
detected in 53 of 58 sight lines, with 59\% of the HVCs showing
negative LSR velocities.  If one sets the \SiIII\ column 
density cutoff at $\log {\rm N}_{\rm SiIII} = 12.50$,
comparable to the detection limit of the poorer data, the covering 
factor drops to $84\pm4$\% (49 of 58 sight lines).  The gas 
appears to be clumpy in some cases, as evidenced by the sight-line pair 
3C~273/Q~1230+0115 (separated by $\lesssim1\arcdeg$) that show very 
different absorption profiles.  In other close pairs, the absorption 
characteristics are similar, indicating extended cloud structure.

\item{\it High-velocity gas appears near the HVC Complexes.} 
A significant number (19 of 58) of the sight lines probe the 
region occupied by Complexes A, C, K, M, and Milky Way Outer Arm.
Even though high-velocity \ion{H}{1}\ 21-cm emission is detected in 
only 9 of the 19 sight lines with ${\rm N}_{\rm HI} \geq1\times10^{19}$ \cd, 
we detect \ion{Si}{3}\ HVCs in all 19 sight lines, in some cases quite 
far ($\sim15\arcdeg$) from any \ion{H}{1}\ emission.  The distribution 
of \ion{Si}{3}\ near the HVC Complexes is somewhat different than that 
of \ion{O}{6}, which is typically not detected more than a few degrees 
from the HVC Complexes.  

\item{\it The \SiIII\ HVC absorber kinematics exhibit a redshift/blueshift 
dipole.} As first seen in our previous studies of HVC kinematics (CSG05), 
we detect a dipole in the pattern of the \ion{Si}{3}\ HVCs split across the 
direction of the Galactic anti-center.  This pattern is seen through a 
segregation of redshifted and blueshifted absorbers across the Galactic 
rotation axis at $\ell = 180^{\circ}$ and is consistent with a lag in 
rotation velocity above the Galactic plane.
Such a pattern can be explained by several models that decouple the
HVCs from rotation of the Galactic disk.

\end{enumerate}

What do these results indicate for the ionization state and total
hydrogen column densities of the HVC population?  With STIS/E140M,
the \SiIII\ $\lambda 1206.50$ absorption line is sensitive to column
densities at or below N$_{\rm SiIII} \approx (10^{12}$~\cd)$N_{12}$,
corresponding to equivalent width $W_{\lambda} =$ (21.0~m\AA)$N_{12}$.
Our best STIS data reach equivalent widths of 10 m\AA, or
$\log N_{\rm SiIII} = 11.7$.  These limits may improve with new
data from the Cosmic Origins Spectrograph (COS) recently installed on 
\HST.  The COS instrument is capable of obtaining far-UV spectra at 
S/N $\geq 40$ and 15 \kms\ resolution, and it should be sensitive to 
\SiIII\ equivalent widths as low as 6 m\AA\ or 
column densities $\log{\rm N}_{\rm SiIII} \approx 11.7$.

As noted above, the \SiIII\ ultraviolet absorption line is a sensitive 
probe of ionized gas, sensitive to total hydrogen column densities 
${\rm N}_{\rm H} \approx 10^{17}$ \cd, for gas at 20\% solar metallicity. 
Analysis of metal ions such as \OVI, \CIV, \SiIII\ in these HVCs currently 
provides our most sensitive probe of low-metallicity gas from 
the Galactic halo -- perhaps the ``cold-mode" accretion that replenishes 
material used up by star formation in the disk (Keres \etal\ 2005;
Dekel \& Birnboim 2006).    
Extending the 21-cm studies down to $\log N_{\rm HI} \approx 17$ would 
better match these UV-line sensitivities and help to sharpen our estimates 
of ionization conditions and mass-inflow rates.  These sensitivities
should be possible with new radio facilities and receivers, including the
ALFA survey with Arecibo (Giovanelli \etal\ 2005), the Greenbank Telescope, 
and various precursors of the Square Kilometer Array.

\acknowledgments

Our group's research support at the University of Colorado for
UV studies of the IGM and Galactic halo gas comes from COS grant
NNX08-AC14G and STScI spectroscopic archive grants (AR-10645.02-A
and AR-11773.01-A).  We also acknowledge theoretical support from 
NSF grant AST07-07474.

\appendix

\section{E140M Sight Lines}

\paragraph{3C 273.}
High-velocity gas toward 3C 273 has been well studied using both {\it
HST}-STIS and {\it FUSE} data (S03, CSG05).  There is a weak
high-velocity \ion{O}{6} absorber in this sight line, 
but no corresponding feature is detectable in any other absorption line,
including \ion{Si}{3}. To establish an upper limit on the
high-velocity \ion{Si}{3} column density, we use the velocity extent
of the high-velocity \ion{O}{6} feature, $V_{\rm LSR}=105$ to $240$ \kms. 

\paragraph{3C 351.}
3C 351 is a Complex C sight line and was investigated 
by Tripp \etal\ (2003) and CSG07.  
The sight line also pierces Complex K and the High-Velocity Ridge at
slightly lower and higher velocities than Complex C, respectively.
These three high-velocity features are resolvable in many absorption
lines.  In \ion{Si}{3}, however, the Complex C and K features are
saturated and blended with lower-velocity absorption.  We therefore
measure these components as one single Complex C/K component and
determine a lower limit to $N_{\rm SiIII}$.

\paragraph{H 1821+643.}
The H 1821+643 sight line passes through the Milky Way Outer Arm and
near Complexes C and K (Tripp \etal\ 2003).  S03 detect high-velocity
\ion{O}{6} all the way out to $V_{\rm LSR}=-285$ km s$^{-1}$.  Lines of
lower ionization stage, however, do not extend to nearly as large a
negative velocity.  In contrast to weaker lines where features of Complex C 
and C/K are well separated, the \ion{Si}{3} is heavily saturated and blended
in this velocity range.  For this reason, we integrate over the full Complex C
and K features for the analysis of the \ion{Si}{3} line.  

\paragraph{HE 0226-4410.}
This sight line passes through a highly ionized HVC that has been
investigated in detail (CSG05, Fox \etal\ 2005).  The HVC is detected
in multiple absorption lines. 

\paragraph{HS 0624+6907.}
This sight line passes through a detectable \ion{H}{1} column of the
Outer Arm.  The Outer Arm feature is clearly detectable in \ion{Si}{3}
over $V_{\rm LSR}=-127$ to $-90$ km s$^{-1}$.  Although this feature is
not detected in \ion{O}{6} (S03), it is detectable in multiple ionization
stages in the E140M bandpass ranging from \ion{O}{1}\ to \ion{C}{4}.  
A weak \ion{Si}{3} feature
($4\sigma$) is also present over $V_{\rm LSR}=100$ to $135$ km s$^{-1}$.
This weak feature is not detected in any other absorption line.
Although we cannot identify it as an IGM feature, the possibility
cannot be ruled out.

\paragraph{HS 1700+6416.}
The HS 1700+6416 sight line passes through a broad weak column of
Complex C.  Based on the \ion{H}{1} (LDS) data toward this sight line,
we cannot measure the \ion{H}{1} column density of Complex C to better
than $3\sigma$, although a feature does seem to be present.  The
Complex C feature is detected in multiple absorption lines, most of
which are saturated, including the \ion{Si}{3} line.

\paragraph{Mrk 279.}
Mrk 279 is an important Complex C sight line, discussed in detail by 
CSG03 and CSG07.  There are two Complex C components
detected in \ion{H}{1} emission, the lower velocity of which blends
with intermediate and low velocity gas.  We take the approach of CSG07
and analyze only the higher-velocity Complex C component.  This
component is detected in multiple absorption lines, with the
\ion{Si}{3} absorption extending to slightly more negative velocity
than the low ions.  We integrate the full \ion{Si}{3} feature for the
column density measurement.

\paragraph{Mrk 335.}
The Mrk 335 sight line shows a complicated \ion{Si}{3} profile with
four possible distinct high-velocity features.  All these features 
are detected in other absorption lines, except the
highest velocity feature, which may have corresponding detections in
other lines at slightly offset velocity.  The broadest
feature centered at $V_{\rm LSR}\approx-325$ km s$^{-1}$ is also detected
in \ion{O}{6} (S03).  The
saturated feature centered at $V_{\rm LSR}\approx-100$ km s$^{-1}$
overlaps in velocity with a detected high-velocity \ion{O}{6} feature.
However, the \ion{O}{6} feature extends to more negative velocity than
the \ion{Si}{3} feature by nearly 100 km s$^{-1}$, suggesting that they 
may not be completely associated with one another.

\paragraph{Mrk 509.}
This sight line is one of the earliest to be examined, defining the category
of highly ionized HVCs (Sembach \etal\ 1999).  CSG04 studied high
velocity gas in this sight line recently and measured ion column
densities for two negative-velocity highly ionized HVC components,   
detectable in ions ranging from \ion{C}{2} through \ion{O}{6}.
The \ion{Si}{3} measurements are updated here, and we also
detect a positive high-velocity wing to the
\ion{Si}{3} profile.  Although it does not extend to as high velocity,
this feature may be related to the \ion{O}{6} wing presented by S03 that 
extends out to $V_{\rm LSR}=200$ \kms.

\paragraph{Mrk 876.}
Mrk 876 is another Complex C sight line investigated by Murphy \etal\ (2000),
CSG03, and CSG07.  There are two Complex C components  pierced by the sight
line to Mrk 876, and they appear saturated and blended together in the
\ion{Si}{3} profile.  The division between the two components is taken
at $V_{\rm LSR}=-155$ km s$^{-1}$, which is the division seen 
in weaker low ionization lines (e.g. \ion{O}{1}\ and \ion{Si}{2}).  

\paragraph{Mrk 1383.}
Although a weak high-velocity positive \ion{O}{6} wing was detected by S03,
we see no evidence of any high-velocity \ion{Si}{3} absorption in this
sight line.  Keeney \etal\ (2006) detected high-velocity absorption in
\OVI, \CIII, \CIV\ and \SiIV.  We estimate an upper limit on
high-velocity $N$(\ion{Si}{3}) by assuming a width of 100 km s$^{-1}$ for 
any possible feature.

\paragraph{NGC 1705.}
This sight line intercepts HVC WW487, which is seen centered at
$V_{\rm LSR}\approx285$ km s$^{-1}$ in both \ion{H}{1} data
(Wakker \& van Woerden 1991) and in multiple
absorption lines in the STIS bandpass
(Vazquez \etal\ 2004).  The \ion{Si}{3} profile shows the WW487
feature and a positive velocity HVC
that extends over $V_{\rm LSR}=100$ to $148$ km s$^{-1}$.  The lower-velocity
HVC is seen
in lines of \ion{Si}{2} and \ion{C}{2} as well.  High-velocity \ion{O}{6} 
associated with both features was detected by S03.
The prominent feature at $V_{\rm LSR}>400$ km s$^{-1}$ is \ion{Si}{3}\ absorption 
intrinsic to NGC 1705.

\paragraph{NGC 3516.}
The NGC 3516 sight line passes near Complex C, but not through a detectable
(21-cm) column of \ion{H}{1}.  A \ion{Si}{3} feature is detected centered at
$V_{\rm LSR}\approx-150$ km s$^{-1}$, a velocity similar to that of
Complex C in that vicinity.  The feature is also clearly detected in
\ion{C}{2} $\lambda1334.53$, and possibly in weak features of 
\ion{O}{1} $\lambda1302.17$ and \ion{Si}{3} $\lambda1193.29$.  This sight line 
likely probes a low column density portion of Complex C.

\paragraph{NGC 3783.}
The \ion{Si}{3} profile shows clear negative velocity high-velocity
gas spread over two components, the higher-velocity component of which
is detectable in \ion{H}{1} emission as HVC WW187 (Wakker \etal\ 2003).  
The high-velocity gas was first noticed in the STIS data by
Gabel \etal\ (2003). The features are clearly seen as two
components in nearly all low ion absorption lines in the E140M
bandpass, but are not seen in \ion{C}{4} or \ion{Si}{4} lines.

\paragraph{NGC 4051.}
No high-velocity gas is detected in the \ion{Si}{3} profile, nor in
any other absorption lines.  We estimate an upper limit on high-velocity 
$N_{\rm SiIII}$ by assuming a width of 100 km s$^{-1}$ for any possible feature.

\paragraph{NGC 4593.} 
High-velocity \ion{Si}{3} is detected in this
sight line in two unusual narrow features, one centered at
$V_{\rm LSR}\approx100$ km s$^{-1}$ and the other at $V_{\rm LSR}\approx275$
km s$^{-1}$.  These detections are confirmed by corresponding
detections of each feature in \ion{Si}{2} $\lambda1260.42$, and of the
lower velocity feature in \ion{C}{2} $\lambda1334.53$.  The lower-velocity 
feature appears to extend out to larger positive velocity in the \ion{Si}{3}\
profile.  Because this part of the 
feature is not present in the \ion{Si}{2}\ and \ion{C}{2}\
profiles, we use the velocity extent of the lower ions for the 
\ion{Si}{3}\ integration.

\paragraph{NGC 5548.}
This sight line shows some evidence for high-velocity gas, but
the absorption only extends 
out to $V_{\rm LSR}=-125$ km s$^{-1}$ in the \ion{Si}{3}
profile.  The feature is not seen in low ions, although low ion
absorption features do extend out to nearly $V_{\rm LSR}=-100$ km
s$^{-1}$.  Although we see absorption out to $V_{\rm LSR}=-125$ km
s$^{-1}$ in \ion{O}{6}, it was reported as a high-velocity
non-detection by S03.  Owing to the weakness of the absorption and
the small extent into the ``high-velocity'' range, we report this sight
line as a \ion{Si}{3} non-detection.  We estimate an upper limit on
high-velocity $N_{\rm SiIII}$ by assuming 100 km s$^{-1}$
width for any possible feature.

\paragraph{NGC 7469.}
Several high-velocity \ion{Si}{3} components are detected in this sight line 
at negative velocity.  Absorption in \ion{O}{6} associated with each of these 
components is reported by S03.  The strong highest-velocity component is
detected in a variety of absorption lines ranging in ionization from
\ion{O}{1} to \ion{C}{4}.  We break this strong absorption feature into 
two components separated at $V_{\rm LSR}=-270$ km s$^{-1}$.  The lowest-velocity
component is not detected in low ions, only in the higher ions 
\ion{C}{4}, \ion{Si}{4}, and \ion{Si}{3}.

\paragraph{PG 0953+414.}
This sight line passes through a positive velocity HVC that was investigated 
in CSG05 and Fox \etal\ (2005).  Both \ion{Si}{3} and \ion{C}{2} $\lambda1334.53$ 
profiles show evidence that this feature consists of two components.
We analyze them in that manner for this work.  We also detect clear high-velocity 
absorption at $V_{\rm LSR}\approx-150$ km s$^{-1}$, where a strong narrow feature is 
detected in \ion{C}{2} and \ion{Si}{2} as well, but not in high ions 
\ion{C}{4} and \ion{O}{6}.  As mentioned by CSG05, this feature correlates 
spatially and kinematically with HVC Complex~M, although it does not pass 
through a detectable \ion{H}{1} column of that object.  Closer to Galactic 
absorption, we detect a negative-velocity wing to the \ion{Si}{3} profile 
extending out to $V_{\rm LSR}=-130$ km s$^{-1}$.  This wing is present in low ions,
\ion{C}{2} and \ion{Si}{2}, as well as \ion{C}{4}.  This feature may be 
associated with Galactic absorption, but since it is technically at
high velocity, we report it as such.

\paragraph{PG 1116+215.}
The positive high-velocity gas in this sight line was first presented in STIS 
data by Sembach \etal\ (2004) and thoroughly studied 
by Ganguly \etal\ (2005) and CSG05.  Although the higher 
velocity component (130--220 \kms) is 
detected in multiple absorption lines ranging in ionization from 
\ion{O}{1} to \ion{C}{4} and \ion{O}{6}, the lower velocity component (90--125
\kms) is present to a high level of certainty 
only in \ion{Si}{3} and \ion{C}{4}.

\paragraph{PG 1211+143}
No high-velocity \ion{O}{6} absorption was detected in this sight line
by the S03 survey.  However, in \ion{Si}{3}, we detect two weak positive 
high-velocity features.  The lower-velocity component is clearly
detected in absorption lines of  \ion{Si}{2} and \ion{C}{2}.  The weak
component at higher velocity is not detected in any other absorption
line.  We are unable to identify it as an IGM line from this
data, nor can it be matched with reported Ly$\alpha$ features in this
sight line reported by Tumlinson \etal\ (2005) and Penton \etal\ (2004; 
based on STIS G140M data).  We identify the feature 
as high-velocity \ion{Si}{3}.

\paragraph{PG 1216+069.}
This sight line shows strong positive high-velocity \ion{Si}{3} absorption 
spread out over at least two components.  This feature is detected in 
\ion{C}{4} over the same velocity extent, and in \ion{Si}{2} $\lambda1260.42$
over a slightly narrower velocity range.  Although not included by S03,
owing to IGM Ly$\beta$ contamination, the feature seems present in
\ion{O}{6} as well.  Based on comparisons of absorption
profiles, we analyze the feature as two components separated at 
$V_{\rm LSR}=234$ km s$^{-1}$.

\paragraph{PG 1259+593.}
This well-studied sight line passes through HVC Complex C (Richter \etal\ 2001;
CSG03; CSG07).  We use the saturated \ion{Si}{3} profile to place a lower 
limit on $N_{\rm SiIII}$ for Complex C in this direction.
A positive-velocity HVC is detected in \ion{O}{6}\ by S03 over 
$V_{\rm LSR}=100-185$ \kms, but no corresponding \ion{Si}{3}\ 
absorption is detected here.

\paragraph{PG 1444+407.}
Possible high-velocity \ion{Si}{3} absorption is seen just blueward of
Galactic absorption in this sight line.  Although the feature is weak,
corresponding high-velocity features are seen in both 
\ion{C}{2} $\lambda1334.53$ and \ion{C}{4} $\lambda1548.20$ over a
similar velocity extent.  No corresponding high-velocity \ion{O}{6}
was detected in the S03 survey.

\paragraph{PHL 1811.}
The complicated high-velocity absorption features in this sight line
were presented in CSG05, but a full quantitative analysis was deferred
until further data was available.  The data presented here include the
full STIS exposure.  We detect strong high-velocity \ion{Si}{3}
absorption in four components spread over a wide velocity range at
negative $V_{\rm LSR}$.  These components are detected in multiple
absorption lines ranging in ionization from \ion{O}{1} to \ion{O}{6}.

\paragraph{PKS 0312-770.}
This sight line passes through the Magellanic Bridge, and strong
corresponding high-velocity absorption is seen in multiple absoprtion
lines, including \ion{Si}{3} $\lambda1206.50$.  Although not apparent
from the \ion{Si}{3}\ line, this feature breaks into two components 
separated at $V_{\rm LSR}=285$ km s$^{-1}$ 
in weaker lines (Shull \etal\ 2009).  
We use the velocity structure seen in the weaker lines
to establish the velocity range for measurement of the two components
in \ion{Si}{3}.  Good \FUSE\ data exists for this sight line, but due
to a break in the spectrum, the \ion{O}{6} lines are unavailable.

\paragraph{PKS 0405-123.}
A weak ($\sim4\sigma$) absorption feature just redward of Galactic
\ion{Si}{3} absorption is detected in this sight line.  We do not
detect absorption at the same velocity in any other absorption lines
in the STIS bandpass.  A weak \ion{O}{6} detection at this velocity was 
presented by S03.  The \FUSE\ exposure time for this target has been
doubled since the work of S03, and we confirm the presence of the
high-velocity \ion{O}{6} feature in those data.  Based on the
corresponding \ion{O}{6} detection and the absence of any known IGM
contaminant lines, we adopt the feature in question as a high-velocity
\ion{Si}{3} detection.  Recent COS observations confirm this HVC, 
with an equivalent width $29 \pm 7$ m\AA.   

\paragraph{PKS 1302-102.}
This sight line contains a strong high-velocity \ion{O}{6} feature (S03, 
CSG05) with no associated absorption in any other ions, including 
\ion{Si}{3}.  We use the $V_{\rm LSR}=200$ to $340$ km s$^{-1}$ extent of 
the \ion{O}{6}\ feature to establish an upper limit
on the \ion{Si}{3}\ absorption.

\paragraph{PKS 2155-304.}
This sight line contains a pair of highly ionized HVCs that were among
the first to be studied using \HST-GHRS data (Sembach \etal\ 1999).
\HST-STIS data for this sight line (Shull, Tumlinson, \& Giroux 2003) was 
analyzed by CSG04, showing absorption in multiple ion stages ranging from 
\ion{C}{2} to \ion{O}{6}.  The absorption is detectable as two distict 
negative-velocity components.

\paragraph{Q 1230+0115.}
Two narrow high-velocity \ion{Si}{3} features are detected in this sight line 
centered at $V_{\rm LSR}\approx125$ and $300$ kms$^{-1}$.  These detections are 
confirmed by associated absorption in multiple lines of \ion{Si}{2}, 
\ion{C}{2}, and \ion{O}{1}.  These features, however, are not detected in the 
higher ions \ion{C}{4} or \ion{Si}{4}.  Wakker \etal\ (2003) presented 
\FUSE\ \ion{O}{6}\ data for this sight line, in which associated absorption 
appears present in the 1031.93 \AA\ line, but not in the 1037.62 \AA\ line.

\paragraph{TON 28.}
High velocity \ion{Si}{3} absoprtion appears over the velocity
range $V_{\rm LSR}=97$ to $183$ km s$^{-1}$.  However, associated
absoprtion over that range is not detected in any other lines in the
\HST-STIS bandpass.  S03 report a detection in \ion{O}{6} over a
nearly identical velocity range, but at low significance
($\sim3\sigma$).  Since we are unable to find any coincident IGM lines
at redshifts that could contaminate the \ion{Si}{3} line, we adopt this as
a high-velocity \ion{Si}{3} detection.

\paragraph{TON S210.}
These data were taken for a program whose goal was to investigate a
compact HVC (CHVC) in this sight line.  Definite high-velocity
\ion{Si}{3} absorption over two slightly blended components is seen,
and \ion{O}{6} absorption over the same velocity extent was presented 
by S03.  Absorption out to $V_{\rm LSR}\approx-150$ km s$^{-1}$ is most likely 
due to an IVC that peaks near
$V_{\rm LSR}\approx-80$ km s$^{-1}$.  The lower velocity component
centered near $V_{\rm LSR}\approx-180$ km s$^{-1}$ is detectable in lines
ranging in ionization from \ion{O}{1} to \ion{O}{6}.  The higher
velocity component centered near $V_{\rm LSR}\approx-240$ km s$^{-1}$ is
seen prominently in the higher ions \ion{C}{4} and \ion{Si}{4}, but
only weakly, if at all, in singly ionized and neutral species.
  
\paragraph{UGC 12163.}
We detect strong high-velocity absorption in this sight line throughout the 
negative velocity range out to nearly $V_{\rm LSR}\approx-500$ km s$^{-1}$.  This 
absorption is seen in ionization stages ranging from \ion{O}{1} to \ion{O}{6}. 
These detections were first reported by CSG05, but the S/N was judged to be 
too poor to properly disentangle the complicated absorption features.  There 
are at least three high-velocity \ion{Si}{3} components present in this sight
line that correspond to features seen in other absorption lines.  Owing to 
the complicated absorption profiles, we rely only on the \ion{Si}{3} profile to
establish integration ranges.

\section{G140M Sight Lines}

\paragraph{ES 438-G009.} 
The \ion{Si}{3} profile shows complicated, yet strong, positive 
high-velocity absorption spread over at least two components.  
Because the spectrum covers only the range 1194--1249 \AA, 
we are unable to confirm the detection with other ion species, nor can we
use other lines to better constrain the velocity ranges.  
We cannot attribute the high-velocity \ion{Si}{3} absorption to any redshifted
contaminant lines, so we treat this as a high-velocity \ion{Si}{3} detection.
According to our analysis of unresolved saturation in binned E140M data 
(see Section~3), the lower velocity component is likely saturated, and we treat 
the measured $N_{\rm SiIII}$ as a lower limit.

\paragraph{HE 0340-2703.}
A narrow feature at positive velocity is detected in the \ion{Si}{3} profile 
for this sight line.  Although the limited coverage of the spectrum 
(1194--1249 \AA\ only) prevents the confirmation of this as a \ion{Si}{3} 
feature from other absorption lines, we cannot attribute this to a redshifted
contaminant line based on the redshifted Ly$\alpha$ features that we can detect
in the spectrum.  We therefore treat this as a high-velocity \ion{Si}{3}
detection.

\paragraph{HE 1029-1401.}
This sight line has excellent data covering the wavelength range
1194--1298 \AA.  A strong positive high-velocity wing is apparent in
the \ion{Si}{3} profile and is confirmed by a nearly identical wing
seen in the profile of \ion{Si}{2} $\lambda1260.42$.  In the weaker
absorption lines of \ion{N}{1} $\lambda1200.71$ and
\ion{S}{2} $\lambda1253.80$, the lower velocity part of the feature is
clearly seen as a single component.  Since there is a small hump also seen on
the shoulder of the \ion{Si}{3} feature, indicating more than one component, 
we break the high-velocity absorption into two components separated 
at $V_{\rm LSR}=178$ \kms.

\paragraph{HS 1543+5921.}
This is a Complex C sight line thoroughly analyzed in CSG07.
The data are of good quality and cover a large wavelength range,
1194--1348 \AA.  In low ions, Complex C can be detected in lines of
\ion{N}{1}, \ion{O}{1}, \ion{Si}{2}, \ion{S}{2}, and \ion{C}{2}.
Corresponding Complex C absorption has a quite different appeareance
in \ion{Si}{3}, lacking any sort of peak near the line centroid.  
A significant absorption feature at $V_{\rm LSR}\approx-400$ \kms\
corresponding to intervening 
\ion{Si}{2} $\lambda1193.29$ at $z=0.0096$ just blueward of the Complex C 
\ion{Si}{3} feature makes continuum fitting near the \ion{Si}{3} line 
difficult.  Thus, the Complex C \ion{Si}{3} profile may be slightly distorted
or contaminated. 

\paragraph{IRAS 08339+6517}
A strong negative high-velocity feature is present in the 
\ion{Si}{3} profile.  The wavelength coverage of the spectrum is limited to 
1194--1249 \AA, so the high-velocity \ion{Si}{3} feature cannot be confirmed 
with other absorption lines.  However the LDS \ion{H}{1}\ 21-cm spectrum shows
a portion of this HVC over $-200<V_{\rm LSR}<-100$ \kms, at a column density
${\rm N}_{\rm HI} \approx2\times10^{19}$ cm$^{-2}$.  Although the \ion{Si}{3}\ 
absorption extends to larger negative velocity, the \ion{H}{1}\ profile confirms
this high-velocity detection.

\paragraph{MCG 10.16.111}
This sight line has excellent quality data, but over a limited velocity range
(1194--1249 \AA).  In \ion{Si}{3}, we see negative high-velocity absorption
extending out to nearly $V_{\rm LSR}=-250$ \kms, which we cannot confirm with other 
absorption lines.  There are numerous Ly$\alpha$
features in this sight line associated with intervening galaxies 
(Bowen, Pettini, \& Blades 2002), but given their strengths in Ly$\alpha$ 
it is highly unlikely that the \ion{Si}{3} profile is contaminated by 
associated absorption in redshifted metal lines.   The \ion{H}{1}\ LDS profile
shows an IVC extending out to $V_{\rm LSR}=-70$ \kms, which may be responsible for 
some of the high-velocity absorption if the feature is strong enough for
saturated wings to blend with the HVC absorption.  However, owing to the 
extension to large negative velocity, we accept this feature as a 
high-velocity \ion{Si}{3} detection.

\paragraph{MRC 2251-178}
Data for this sight line are of excellent quality and cover the
wavelength range 1194--1298 \AA.  The sight line intercepts a 
negative-velocity highly-ionized HVC analyzed in detail by CSG05.  
In the \FUSE\ bandpass, \ion{O}{6} and \ion{C}{2}~$\lambda1036.34$ are
detected over the same velocity extent as the negative-velocity
\ion{Si}{3} HVC.  In addition, we detect a positive-velocity wing to
the \ion{Si}{3} profile that extends out beyond $V_{\rm LSR}=150$ \kms\
that is not detected at a statistically significant level in any other
absorption line.  The feature seems present at a very low level in the
\ion{Si}{2} $\lambda1260.42$ line, lending some confidence that the 
\ion{Si}{3} feature is real.

\paragraph{Mrk 110}
A strong negative high-velocity feature is present in the \ion{Si}{3}
profile.  We are unable to confirm the detection with other absorption
lines owing to the limited 1194--1249 \AA\ coverage of the spectrum.
Since we cannot identify the feature as a higher-redshift contaminant, we 
adopt the feature as a high-velocity \ion{Si}{3} detection.  Our analysis of 
saturation in binned E140M datasets indicates that saturation in this strong
feature is likely, and we therefore adopt the measured $N_{\rm SiIII}$
as a lower limit.

\paragraph{Mrk 478}
A weak negative high-velocity wing is present in the \ion{Si}{3}
profile for this sight line out to $V_{\rm LSR}=-158$ \kms.  The 1194--1298
\AA\ coverage of this spectrum allows the analysis of this feature in
additional lines.  In \ion{Si}{2} $\lambda1260.42$, this wing can be
detected out to $V_{\rm LSR}=-135$ \kms\ at a low but statistically
significant level.  S03 report an \ion{O}{6} HVC in this sight 
line over $340<V_{\rm LSR}<435$ \kms.  We do not detect a corresponding feature
in the \ion{Si}{3} line.

\paragraph{Mrk 926}
Clear negative high-velocity absorption is present in the \ion{Si}{3}
profile for this sight line.  This detection was reported in the
highly ionized HVC study of CSG05, but was dropped from thorough
analysis owing to the poor quality of its \FUSE\ data.  Although the STIS
spectrum covers the range 1194--1298 \AA\, we cannot confirm the \ion{Si}{3}
detection since the negative high-velocity range of the strong
\ion{Si}{2} $\lambda1260.42$ line is contaminated by Galactic
\ion{S}{2} $\lambda1259.52$.  However, the poor-quality \FUSE\ data do
provide some evidence of associated absorption features in both
\ion{O}{6} and \ion{C}{2} $\lambda1036.34$.  In addition, there are no 
known higher-redshift contaminants that could be responsible for the feature 
in the \ion{Si}{3} profile.  We therefore adopt this as a high-velocity 
\ion{Si}{3} detection.

\paragraph{Mrk 1044}
Good quality data exist for this sight line from 1194--1298 \AA\ 
and 1540--1594 \AA, which allows the analysis of the \ion{C}{4} doublet.  
An obvious negative high-velocity feature is
present in the \ion{Si}{3} profile.  This feature, however, is not
present in the \ion{C}{4} $\lambda1548.20$ line.  In addition, we
cannot confirm the detection with the \ion{Si}{2} $\lambda1260.42$ line
due to contamination from Galactic 
\ion{S}{2} $\lambda1259.52$.  However, a very
similar absorption feature is seen towards NGC 985, which is
only $1.1\arcdeg$ away from this sight line.  Since there are no known
higher-redshift contaminants that could affect the \ion{Si}{3}
profile, we adopt it as a high-velocity \ion{Si}{3} detection.

\paragraph{Mrk 1513}
This sight line contains highly ionized negative high-velocity gas 
studied thoroughly by CSG05.  The feature is clearly present in the
\ion{Si}{3} line, as well as the \ion{O}{6} and \ion{C}{3} $\lambda977.02$
lines in \FUSE\ data of fair quality.  It is not present at a statistically 
significant level in \ion{C}{2} $\lambda1036.34$.  The STIS data quality is 
superior to the \FUSE\ data, and although the spectrum covers 1194--1298 \AA, 
we cannot confirm the feature in \ion{Si}{2} $\lambda1260.42$ due 
to contamination from Galactic \ion{S}{2} $\lambda1259.52$.  The \ion{Si}{3} profile 
indicates multi-component absorption, and we thus break the feature into
three separate components.

\paragraph{NGC 985}
High-velocity \ion{Si}{3} is present in this sight line spread over
two negative velocity components.  The spectrum covers 1194--1298 \AA,
allowing analysis of the \ion{Si}{2} $\lambda1260.42$ line.
Most of the corresponding velocity range of the prospective HVC is
blocked by Galactic \ion{S}{2} $\lambda1259.52$, but the lower
velocity component is clearly present.  In addition, a similar absorption
feature is seen towards Mrk 1044, which is $1.1\arcdeg$ from this sight line.
Since there are also no known higher redshift contaminants that could affect 
the \ion{Si}{3} profile, the two components are thus accepted as
high-velocity \ion{Si}{3} detections.

\paragraph{PG 0804+761}
High quality STIS data for this sight line are available (Richter \etal\ 2001)
over the wavelength range 1194-1298 \AA.  The \ion{H}{1} LDS profile for this
sight line shows three components: a Galactic feature, an IVC over
$-80<V_{\rm LSR}<-40$ \kms, and broad low-level high velocity absorption
beginning at $V_{\rm LSR}=-80$ \kms\ and extending out to
$V_{\rm LSR}\approx-170$ \kms.  These three features are well blended in
the \ion{Si}{3} profile, and we use the $V_{\rm LSR}=-80$ \kms\ point to
mark the cutoff for integration of the high-velocity feature.  
The high-velocity feature is also present in the profile of 
\ion{Si}{2} $\lambda1260.42$ out to $V_{\rm LSR}=-150$ \kms, where the feature
becomes blended with \ion{S}{2} $\lambda1259.52$.  This feature was not 
detected in the survey of high-velocity \ion{O}{6} (S03).

\paragraph{PG 1049-005}
The \ion{Si}{3} profile shows strong positive high-velocity absorption, 
spread over two components.  Although the wavelength coverage is limited 
(1194--1249 \AA), the lower velocity HVC component is so strong that it can 
be detected in the weak \ion{N}{1} $\lambda1200.71$ line, confirming the 
high-velocity \ion{Si}{3} detection.  The higher velocity HVC component can 
only be detected in \ion{Si}{3}, so its detection as high-velocity
\ion{Si}{3} cannot be confirmed.  We note that the relatively high
redshift of this QSO target ($z=0.3599$, one of the highest in this
survey) allows for the larger likelihood of contamination by
IGM absorbers.  The limited coverage of the spectrum
prevents a thorough investigation of this possibility.

\paragraph{PG 1351+640}
This sight line pierces HVC Complex C and has been discussed in both
CSG03 and CSG07.  In the \FUSE\ bandpass, this high-velocity feature
can be detected in lines of \ion{Fe}{2}, \ion{Si}{2}, and \ion{O}{6}.
Complex C is highly saturated in the \ion{Si}{3} line and we take its 
measurement as a lower limit on $N$(\ion{Si}{3}).

\paragraph{PKS 1004+130}
The \ion{Si}{3} profile of this sight line shows a strong positive
high-velocity feature.  Owing to the limited 1194--1249 \AA\ coverage
of the spectrum, we cannot confirm its detection with other absorption
lines.  Based on supplementary \FUSE\ and GHRS/G140L data, we cannot
identify the feature in the \ion{Si}{3} profile as a contaminant.  We
therefore adopt it as a high-velocity \ion{Si}{3} detection.  
This sight line was reported as an \ion{O}{6} non-detection by S03.  
However, the \ion{O}{6} $\lambda1031.926$ profile shows an absorption 
feature with an identical velocity extent and trough as the \ion{Si}{3} 
feature, although with a slightly different overall appearance.  This 
feature is not present in \ion{O}{6} $\lambda1037.62$, although that part 
of the spectrum is strongly contaminated by H$_{2}$ lines, leading to the 
S03 classification as an \ion{O}{6} non-detection.  Wakker \etal\ (2003)
suggest that the feature in the \ion{O}{6}\ profile may be
\ion{O}{3} $\lambda832.93$ at $z=0.2398$.  The detection of corresponding
high-velocity \ion{Si}{3} opens up the possibility that the feature
in the \FUSE\ spectrum may actually be high-velocity \ion{O}{6}.

\paragraph{PKS 1103-006}
The data quality for this sight line is sub-par, but sufficient
to detect the strong positive high-velocity absorption in the
\ion{Si}{3} profile.  Owing to the limited 1194--1249 \AA\ wavelength
coverage, we are unable to confirm this detection with other
absorption lines.  The relatively high redshift of this target
($z=0.423$) increases the likelihood of contamination by IGM lines, but from 
the available data we cannot identify a definite contaminant.  Thus, we 
accept the feature as a detection of high-velocity \ion{Si}{3}.

\paragraph{PKS 2005-489}
A strong positive high-velocity wing to Galactic absorption is seen in the 
\ion{Si}{3}\ profile for this sight line.  The data quality is excellent,
with coverage extending from 1194--1298 \AA.  In addition to this HVC being
detected in \ion{O}{6}\ (S03), it can also be detected as a wing in
\ion{Si}{2} $\lambda1260.42$ and \ion{N}{1} $\lambda1200.71$.
Keeney \etal\ (2006) also reported high-velocity \OVI\ and \CIII\
(the \CIV\ and \SiIV\ spectra were not available in STIS/E140M spectra,]
owing to the 2004 failure.) 
 
\paragraph{Q 1831+731}
The data quality for this sight line is good, but the wavelength
coverage is limited to the 1194--1249 \AA\ range.  A fairly strong
negative high-velocity wing to Galactic absorption is seen in the
\ion{Si}{3}\ profile extending out to $V_{\rm LSR}\approx-200$ \kms.
Given the limited wavelength coverage, we cannot conclusively confirm
this feature with other absorption lines.  The \ion{N}{1} $\lambda1199.55$ 
line shows a negative velocity wing extending out to possibly
$V_{\rm LSR}\approx-150$ \kms, although not at a level of sufficient 
significance.  Since we cannot identify a higher-redshift contaminant 
that could be affecting the \ion{Si}{3} profile, we accept this as a 
high velocity \ion{Si}{3} detection.

\paragraph{RXSJ 01005-511}
Strong positive high-velocity absorption is seen over possibly two
components in the \ion{Si}{3}\ profile.  The wavelength coverage of
the spectrum, 1194--1298 \AA\ allows access to additional absorption
lines such as \ion{Si}{2} $\lambda1260.422$.  In this \ion{Si}{2}\
line, the two-component structure is more pronounced, with the feature
broken up into a positive-velocity wing and a distinct component that 
peaks at $V_{\rm LSR}\approx175$ \kms.  The feature is not present in
\ion{N}{1} $\lambda1200.71$.  We use the component structure in the 
\ion{Si}{2} profile for the velocity ranges used to measure \ion{Si}{3}.

\paragraph{TON 730}
A weak negative high-velocity wing extending out to $V_{LSR}=-140$
\kms\ can be detected in the \ion{Si}{3}\ profile at the $3\sigma$
level in this sight line.  Owing to the limited wavelength coverage,
we cannot confirm the detection with other absorption lines.  Since
there are no known higher-redshift contaminants that could be
responsible for the absorption, we adopt it as a high-velocity
\ion{Si}{3}\ detection.  However, the fact that it does not extend to
velocities far beyond Galactic absorption makes this a somewhat
questionable detection.  Given the uncertainties in calibrating the
LSR velocity scales of these particular spectra, it is possible that 
the wing lies at a lower velocity and is not high-velocity gas at all.

\paragraph{TON 1542}
A negative high-velocity feature is present in the \ion{Si}{3} profile
at a low level.  The wavelength coverage of the spectrum is fairly wide, 
1194--1298 \AA\, but the velocity of this possible HVC is badly blended
in the \ion{Si}{2} $\lambda1260.42$ profile by Galactic 
\ion{S}{2} $\lambda1259.52$.  Thus we are unable to confirm this detection.  
Since we cannot identify any contaminants at higher redshift 
responsible for the feature, we accept it as high-velocity \ion{Si}{3}\
detection.  However, this detection is at low significance ($\sim3\sigma$).

\paragraph{TON S180}
The data for this target are excellent and cover 1194--1298 \AA.
There is a strong negative highly-ionized HVC in this sight line
thoroughly investigated by CSG05.  In the \FUSE\ bandpass, the 
feature was reported as an \ion{O}{6}\ detection (S03), and it can also be
detected in \ion{C}{2} $\lambda1036.34$ and \ion{C}{3} $\lambda977.02$.
In the STIS spectrum, a corresponding feature is possibly present in 
\ion{Si}{2} $\lambda1260.42$, but most of the feature is severely blended
with Galactic \ion{S}{2} $\lambda1259.52$.

\paragraph{VIIZW118}
Negative high-velocity absorption is seen in the \ion{Si}{3}\ profile for this 
sight line.  The data quality is fair, but the limited wavelength coverage 
of 1194--1249 \AA\ prevents the confirmation of this detection.  In the \FUSE\ 
bandpass, this was reported as an \ion{O}{6}\ non-detection (S03).  A definite
high-velocity feature is seen in \ion{C}{2} $\lambda1036.34$ over a similar
velocity range as the \ion{Si}{3}\ feature.  Since there are no known  
IGM contaminants that could affect the \ion{Si}{3} profile, we adopt it as a 
high-velocity \ion{Si}{3}\ detection.

\clearpage

\begin{deluxetable}{lcccccc}
\tablecolumns{7}
\tablewidth{0pc}
\tablecaption{High-Velocity \ion{Si}{3} Detections from STIS E140M datasets}
\tablehead{
\colhead{Sight Line} & \colhead{$l$} & \colhead{$b$} & \colhead{$v_{min}$} & 
\colhead{$v_{max}$} &  \colhead{$W_{\lambda}$} & \colhead{log$N$(\ion{Si}{3})}\\
\colhead{} & \colhead{(deg)} & \colhead{(deg)} & \colhead{(km s$^{-1}$)} & 
\colhead{(km s$^{-1}$)} & \colhead{(m\AA)} & \colhead{($N$ in cm$^{-2}$)}}
\startdata
3C 351       &  90.09 &  36.38 & -221 & -150 & 195$\pm$11 & $>13.37$\tablenotemark{a} \\
             &        &        & -150 &  -87 & 191$\pm$9  & $>13.40$ \\ 
H 1821+643   &  94.00 &  27.41 & -165 &  -75 & 304$\pm$6  & $>13.70$ \\ 
HE 0226-4410 & 253.94 & -65.77 &  119 &  215 & 233$\pm$14 & $>13.27$\tablenotemark{a} \\
HS 0624+6907 & 145.71 &  23.35 & -127 &  -90 & 111$\pm$7  & $>13.06$\tablenotemark{a} \\
             &        &        &  105 &  135 &  28$\pm$7  & $12.22^{+0.10}_{-0.13}$  \\
HS 1700+6416 &  94.40 &  36.16 & -197 &  -85 & 378$\pm$33 & $>13.75$                 \\ 
Mrk 279      & 115.04 &  46.86 & -226 & -115 & 306$\pm$7  & $>13.55$\tablenotemark{a} \\ 
Mrk 335      & 108.77 & -41.43 & -426 & -380 &  41$\pm$10 & $12.35^{+0.10}_{-0.13}$  \\
             &        &        & -360 & -275 & 125$\pm$14 & $12.90^{+0.05}_{-0.05}$  \\
             &        &        & -257 & -237 &  25$\pm$5  & $12.17^{+0.09}_{-0.11}$  \\
             &        &        & -132 &  -80 & 137$\pm$9  & $>13.20$        \\ 
Mrk 509      &  35.97 & -29.86 & -338 & -256 & 191$\pm$12 & $>13.28$\tablenotemark{a} \\ 
             &        &        & -256 & -213 &  48$\pm$8  & $12.45^{+0.07}_{-0.08}$  \\
             &        &        & 100  &  147 &  80$\pm$14  & $12.74^{+0.08}_{-0.11}$  \\
Mrk 876      &  98.27 &  40.38 & -210 & -155 & 178$\pm$6  & $>13.37$   \\
             &        &        & -155 & -100 & 212$\pm$6  & $>13.56$   \\ 
NGC 1705     & 261.08 & -38.74 & 100  &  148 &  55$\pm$6  & $12.52^{+0.05}_{-0.06}$  \\
             &        &        & 220  &  367 & 322$\pm$13 & $>13.56$\tablenotemark{a} \\ 
NGC 3516     & 133.24 &  42.40 & -187 & -115 & 162$\pm$18 & $>13.13$\tablenotemark{a}  \\
NGC 3783     & 287.46 &  22.95 & 100 &  153 & 113$\pm$17  & $12.98^{+0.07}_{-0.09}$  \\
             &        &        &  153 &  203 &  63$\pm$3  & $12.58^{+0.02}_{-0.02}$  \\
             &        &        &  203 &  288 & 117$\pm$4  & $12.87^{+0.01}_{-0.01}$  \\
NGC 4593     & 297.48 &  57.40 &  88 & 118  &  43$\pm$6  & $12.43^{+0.06}_{-0.08}$  \\ 
             &        &        & 258  & 291  &  36$\pm$7  & $12.33^{+0.08}_{-0.10}$  \\
NGC 7469     &  83.10 & -45.47 & -410 & -270 & 405$\pm$27 & $>13.70$                 \\
             &        &        & -270 & -237 &  70$\pm$9  & $>12.71$\tablenotemark{a}  \\
             &        &        & -190 & -114 &  83$\pm$19 & $>12.68$\tablenotemark{a}  \\
PG 0953+414  & 179.79 &  51.71 & -171 & -130 &  64$\pm$6  & $12.63^{+0.04}_{-0.04}$  \\
             &        &        & -130 & -100 &  78$\pm$20 & $>12.99$\tablenotemark{a} \\
             &        &        &   85 &  165 & 101$\pm$9  & $12.79^{+0.03}_{-0.04}$  \\
             &        &        &  165 &  194 &  20$\pm$5  & $12.05^{+0.09}_{-0.11}$  \\
PG 1116+215  & 223.36 &  68.21 &  90 &  125 &  63$\pm$4  & $12.66^{+0.03}_{-0.03}$  \\
             &        &        &  130 &  220 & 211$\pm$7  & $>13.36$\tablenotemark{a} \\
PG 1211+143  & 267.55 &  74.32 &  160 &  208 &  49$\pm$4  & $12.45^{+0.03}_{-0.03}$  \\
             &        &        &  265 &  293 &  22$\pm$3  & $12.08^{+0.05}_{-0.06}$  \\
PG 1216+069  & 281.07 &  68.14 &  165 &  234 & 110$\pm$14 & $12.87^{+0.05}_{-0.06}$  \\
             &        &        &  234 &  302 & 134$\pm$13 & $>13.08$\tablenotemark{a} \\
PG 1259+593  & 120.56 &  58.05 & -180 & -105 & 184$\pm$7  & $>13.39$\tablenotemark{a}  \\
PG 1444+407  &  69.90 &  62.72 & -164 & -109 &  65$\pm$11 & $12.57^{+0.07}_{-0.08}$  \\
PHL 1811     &  47.48 & -44.82 & -385 & -328 &  46$\pm$6  & $12.44^{+0.05}_{-0.05}$  \\
             &        &        & -309 & -237 & 138$\pm$6  & $>13.03$\tablenotemark{a} \\
             &        &        & -237 & -180 & 187$\pm$5  & $>13.31$ \\
             &        &        & -180 & -128 & 124$\pm$5  & $>13.20$ \\
PKS 0312-770 & 293.44 & -37.55 &  70  &  280 & 673$\pm$30 & $>13.96$ \\
             &        &        & 280  &  372 & 235$\pm$15 & $>13.39$\tablenotemark{a} \\
PKS 0405-123 & 204.93 & -41.76 & 100  &  161 &  43$\pm$10 & $12.39^{+0.08}_{-0.10}$  \\
PKS 2155-304 &  17.73 & -52.25 & -309 & -215 &  34$\pm$8  & $12.25^{+0.09}_{-0.11}$  \\
             &        &        & -195 & -91  & 170$\pm$8  & $>13.16$\tablenotemark{a} \\
Q 1230+0115  & 291.26 &  63.66 &  95  &  152 &  54$\pm$8  & $12.55^{+0.05}_{-0.07}$  \\
             &        &        &  285 &  318 &  49$\pm$5  & $12.51^{+0.05}_{-0.05}$  \\
TON 28       & 200.09 &  53.21 &  97  &  183 &  89$\pm$18 & $12.71^{+0.07}_{-0.10}$  \\
TON S210     & 224.98 & -83.16 & -266 & -221 &  74$\pm$8  & $>12.78$\tablenotemark{a} \\
             &        &        & -221 & -148 & 205$\pm$10 & $>13.41$    \\
UGC 12163    &  92.14 & -25.34 & -477 & -381 & 293$\pm$30 & $>13.54$     \\
             &        &        & -372 & -324 & 104$\pm$19 & $>12.88$\tablenotemark{a} \\
             &        &        & -172 & -135 &  69$\pm$16 & $>12.72$\tablenotemark{a} \\
\enddata
\tablenotetext{a}{Although not apparent from the binned data 
presented in Figure 1, saturation is present in the unbinned data for 
this sight line.  The measured column density is thus taken as a lower limit.} 
\end{deluxetable}

\clearpage

\begin{deluxetable}{lcccccc}
\tablecolumns{7}
\tablewidth{0pc}
\tablecaption{High-Velocity \ion{Si}{3} Detections from STIS G140M datasets}
\tablehead{
\colhead{Sight Line} & \colhead{$l$} & \colhead{$b$} & \colhead{$v_{min}$} & 
\colhead{$v_{max}$} &  \colhead{$W_{\lambda}$} & \colhead{log$N$(\ion{Si}{3})}\\
\colhead{} & \colhead{(deg)} & \colhead{(deg)} & \colhead{(km s$^{-1}$)} & 
\colhead{(km s$^{-1}$)} & \colhead{(m\AA)} & \colhead{($N$ in cm$^{-2}$)}}
\startdata
ES 438-G009  & 277.55 &  29.36 &   65 &  168 & 294$\pm$40 & $>13.45$\tablenotemark{a} \\
             &        &        &  168 &  268 & 123$\pm$46 & $12.87^{+0.14}_{-0.19}$  \\ 
HE 0340-2703 & 222.68 & -52.13 &  275 &  302 &  55$\pm$8  & $12.57^{+0.07}_{-0.09}$  \\ 
HE 1029-1401 & 259.34 &  36.52 &  92 &  178 & 243$\pm$6  & $13.34^{+0.01}_{-0.01}$ \\
             &        &        &  178 &  263 & 130$\pm$7  & $12.91^{+0.02}_{-0.03}$  \\
HS 1543+5921 &  92.40 &  46.36 & -189 & -109 & 88$\pm$32  & $12.72^{+0.13}_{-0.20}$  \\ 
IRAS 08339+6517 & 150.46 & 35.60 & -273 & -100 & 335$\pm$57 & $13.37^{+0.06}_{-0.08}$ \\ 
MCG 10.16.111 & 144.22 & 55.08 & -235 & -100 & 221$\pm$18 & $13.29^{+0.05}_{-0.06}$ \\ 
MRC 2251-178 &  46.20 & -61.33 & -324 & -179 & 145$\pm$13 & $12.94^{+0.03}_{-0.04}$  \\
             &        &        &  100 &  154 &  54$\pm$10 & $12.50^{+0.08}_{-0.12}$  \\
Mrk 110      & 165.01 &  44.36 & -177 & -100 & 196$\pm$24 & $>13.20$\tablenotemark{a} \\ 
Mrk 478      &  59.24 &  65.04 & -158 & -100 &  35$\pm$8  & $12.27^{+0.09}_{-0.11}$  \\
Mrk 926      &  64.09 & -58.76 & -329 & -130 & 221$\pm$61 & $13.10^{+0.10}_{-0.13}$  \\ 
Mrk 1044     & 179.70 & -60.48 & -252 & -160 &  79$\pm$17 & $12.63^{+0.08}_{-0.11}$  \\ 
Mrk 1513     &  63.67 & -29.07 & -340 & -241 &  97$\pm$11 & $12.74^{+0.04}_{-0.05}$  \\
             &        &        & -241 & -160 &  60$\pm$9  & $12.51^{+0.06}_{-0.07}$  \\
             &        &        & -160 & -109 &  23$\pm$7  & $12.06^{+0.11}_{-0.15}$  \\
NGC 985      & 180.84 & -59.49 & -262 & -175 &  51$\pm$10 & $12.42^{+0.08}_{-0.09}$  \\
             &        &        & -175 & -120 &  29$\pm$7  & $12.18^{+0.10}_{-0.12}$  \\ 
PG 0804+761  & 138.28 &  31.03 & -176 & -80  & 143$\pm$16 & $12.99^{+0.06}_{-0.07}$  \\
PG 1049-005  & 252.28 &  49.88 &   85 &  205 & 369$\pm$39 & $>13.55$\tablenotemark{a} \\
             &        &        &  205 &  276 & 117$\pm$30 & $12.89^{+0.10}_{-0.13}$  \\
PG 1351+640  & 111.89 &  52.02 & -213 & -100 & 341$\pm$11 & $>13.70$      \\ 
PKS 1004+130 & 225.12 &  49.12 &  125 &  357 & 334$\pm$52 & $>13.36$\tablenotemark{a} \\
PKS 1103-006 & 256.66 &  52.30 &  115 &  295 & 506$\pm$84 & $>13.72$     \\
PKS 2005-489 & 350.37 & -32.60 &  100 &  232 & 156$\pm$15  & $13.01^{+0.06}_{-0.07}$  \\
Q 1831+731   & 104.04 &  27.40 & -202 & -100 & 141$\pm$17  & $13.02^{+0.04}_{-0.05}$ \\
RXSJ 01004-511 & 299.49 & -65.84 & 100 &  160 & 159$\pm$16 & $13.10^{+0.05}_{-0.05}$ \\
             &        &        &  160 &  257 & 148$\pm$14 & $12.98^{+0.04}_{-0.04}$  \\
TON 730      &  28.71 &  78.15 & -140 & -100 &  33$\pm$11 & $12.26^{+0.12}_{-0.17}$  \\
TON 1542     & 269.44 &  81.74 & -180 & -100 &  37$\pm$12 & $12.28^{+0.12}_{-0.17}$  \\
TON S180     & 139.00 & -85.07 & -232 & -70  & 289$\pm$10 & $13.29^{+0.02}_{-0.01}$  \\
VIIZW118     & 151.36 &  25.99 & -189 & -100 &  81$\pm$15 & $12.66^{+0.07}_{-0.08}$  \\
\enddata
\tablenotetext{a}{Although not apparent from the data 
presented in Figure 2, unresolved saturation is likely present based on
the analysis method discussed in \S\ 3.  
The measured column density is thus taken as a lower limit.} 
\end{deluxetable}

\clearpage

\begin{deluxetable}{lccccc}
\tablecolumns{6}
\tablewidth{0pc}
\tablecaption{Sight Lines With No Detected High-Velocity \ion{Si}{3}}
\tablehead{
\colhead{Sight Line} & \colhead{$l$} & \colhead{$b$} &  \colhead{Velocity}  &
    \colhead{$W_{\lambda}$} & \colhead{log$N$(\ion{Si}{3})}\\
\colhead{} & \colhead{(deg)} & \colhead{(deg)} & \colhead{Range (km~s$^{-1}$)} &  
\colhead{(m\AA)} & \colhead{($N$ in cm$^{-2}$)}}
\startdata
3C 273       & 289.95 &  64.36 & 105 to 240       & $<16$  &  $<11.89$  \\
Mrk 1383     & 349.22 &  55.13 & 100 to200        & $<38$  &  $<12.26$  \\        
NGC 4051     & 148.89 &  70.09 & $-250$ to $-150$ & $<62$  &  $<12.49$  \\ 
NGC 5548     &  31.96 &  70.50 & $-150$ to $-100$ & $<36$  &  $<12.24$  \\ 
PKS 1302-102 & 308.59 &  52.16 & 220 to 340       & $<60$  &  $<12.46$  \\ 

\enddata
\tablecomments{Upper limits ($3 \sigma$) over stated velocity range, 
using E140M echelle data toward 5 sight lines with no high-velocity \ion{Si}{3}. }   
\end{deluxetable}

\newpage

\begin{figure}
\figurenum{1}
\plotone{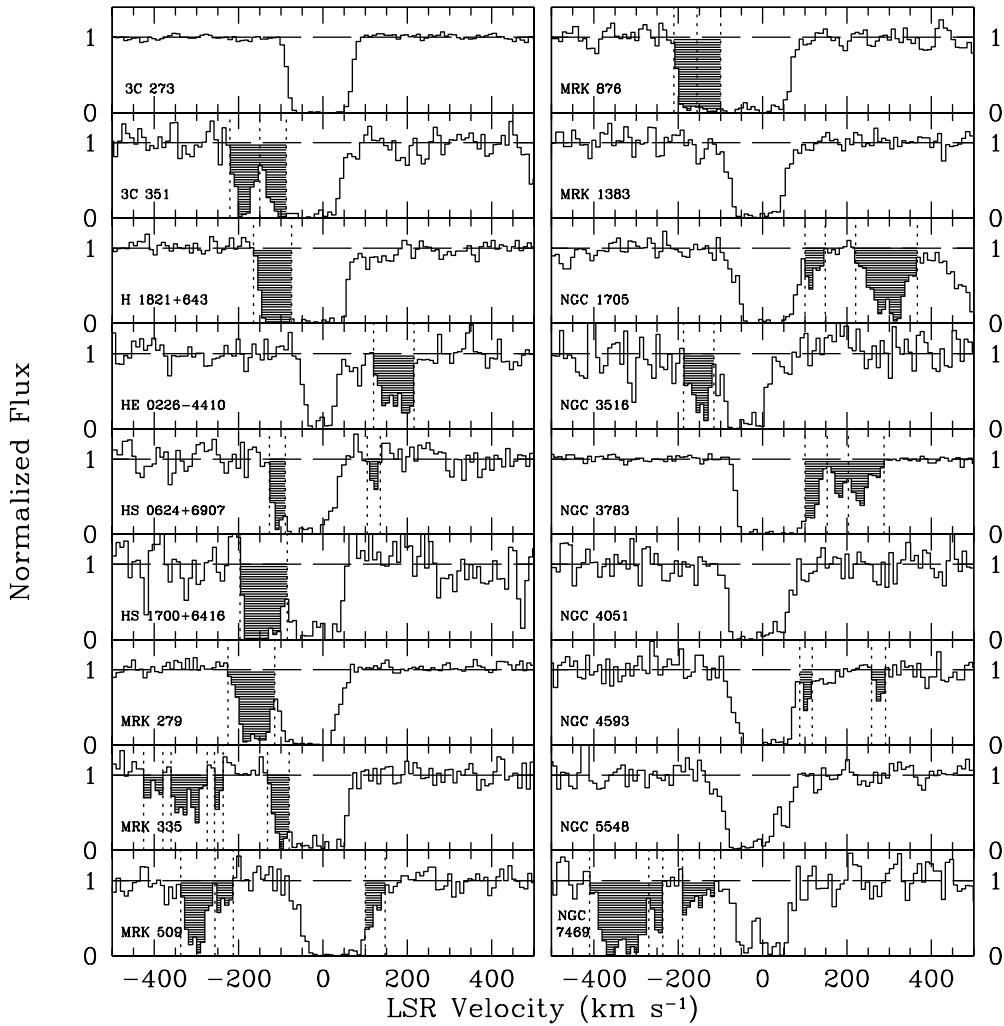}
\caption{Normalized profiles of the \ion{Si}{3} $\lambda1206.50$ line
for 33 sight lines observed with the STIS/E140M echelle.  Dashed lines
indicate velocity extent of detected HVC components;  HVC absorption
has been shaded. }
\end{figure} 

\begin{figure}
\figurenum{1}
\plotone{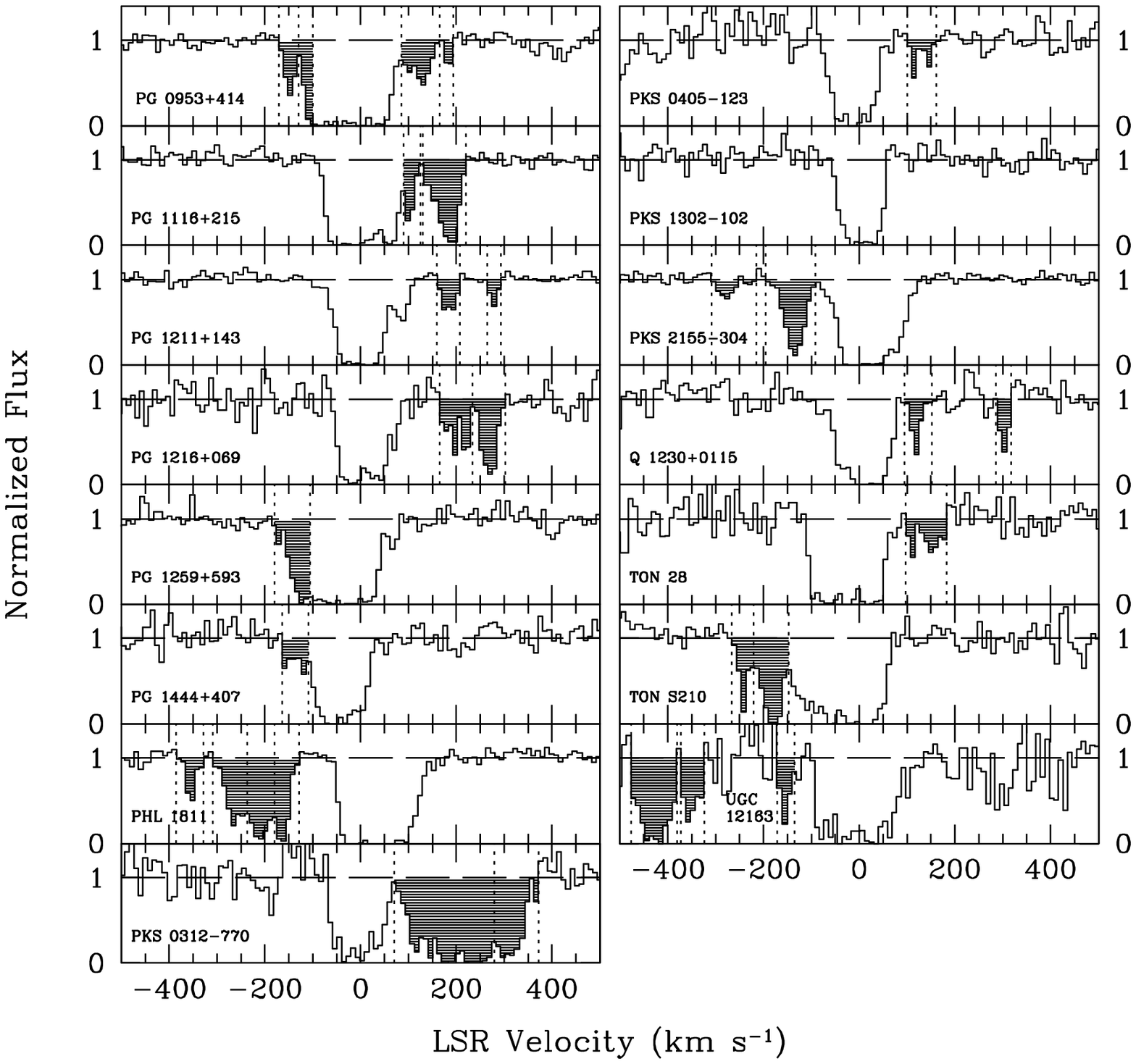}
\caption{Normalized profiles of the \ion{Si}{3} $\lambda1206.50$ line
for 33 sight lines observed with the STIS/E140M echelle.}
\end{figure} 

\begin{figure}
\figurenum{2}
\plotone{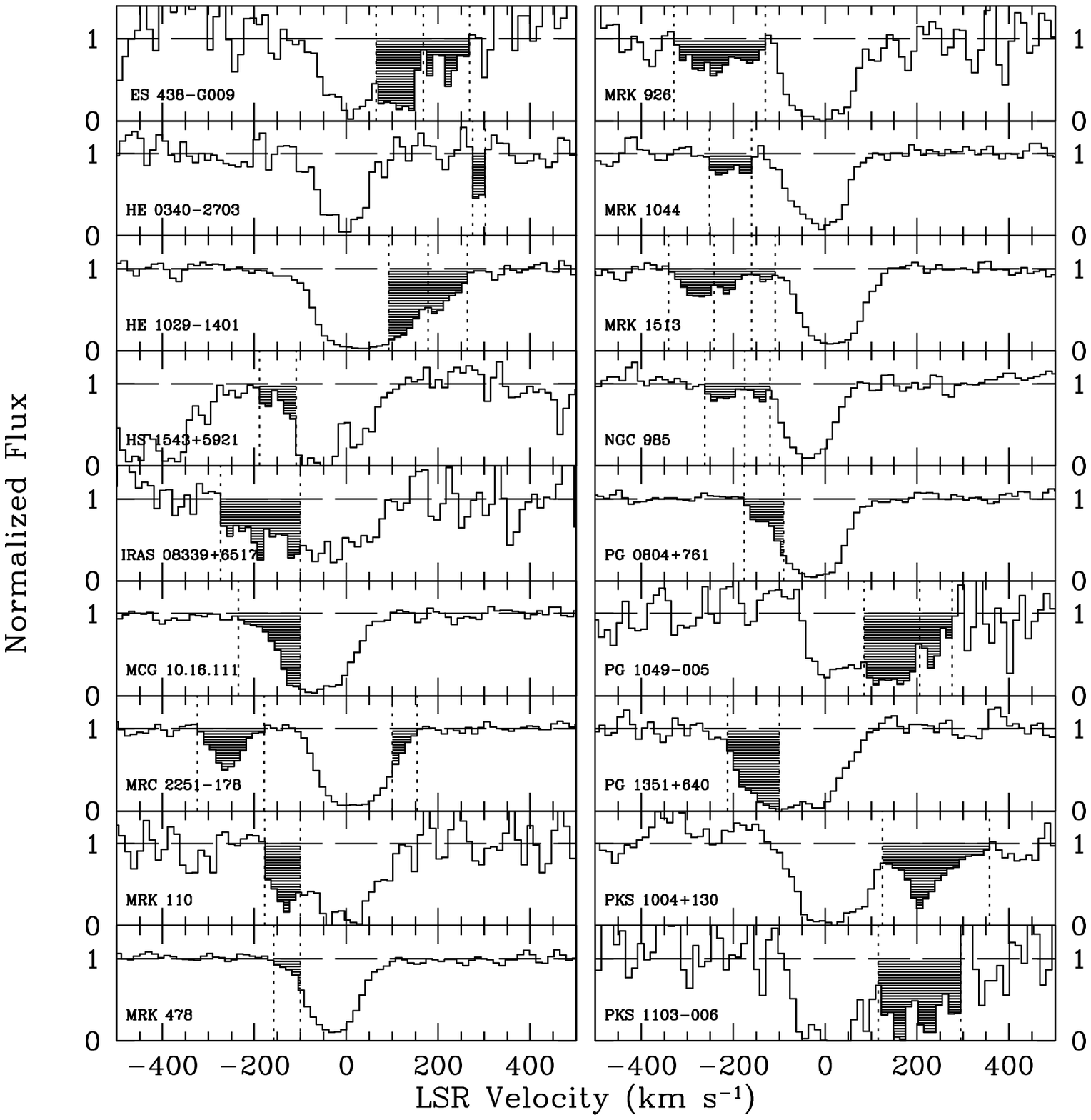}
\caption{Same as Figure 1. 
Normalized profiles of the \ion{Si}{3} $\lambda1206.50$ line
for 25 sight lines observed with the STIS/G140M grating.}
\end{figure} 

\begin{figure}
\figurenum{2}
\plotone{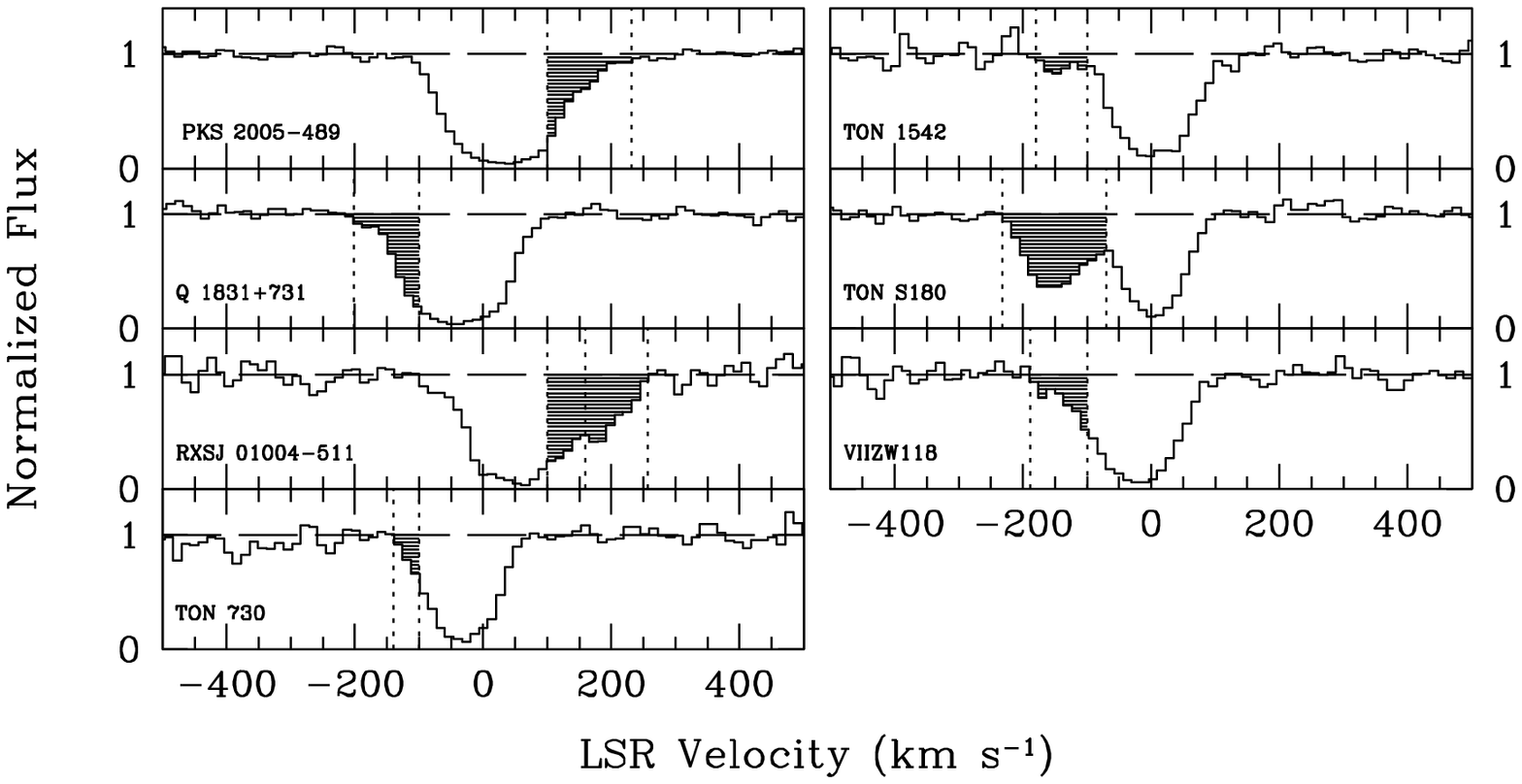}
\caption{Normalized profiles of the \ion{Si}{3} $\lambda1206.50$ line
for 25 sight lines observed with the STIS/G140M grating.}
\end{figure} 

\begin{figure}
\figurenum{3}
\plotone{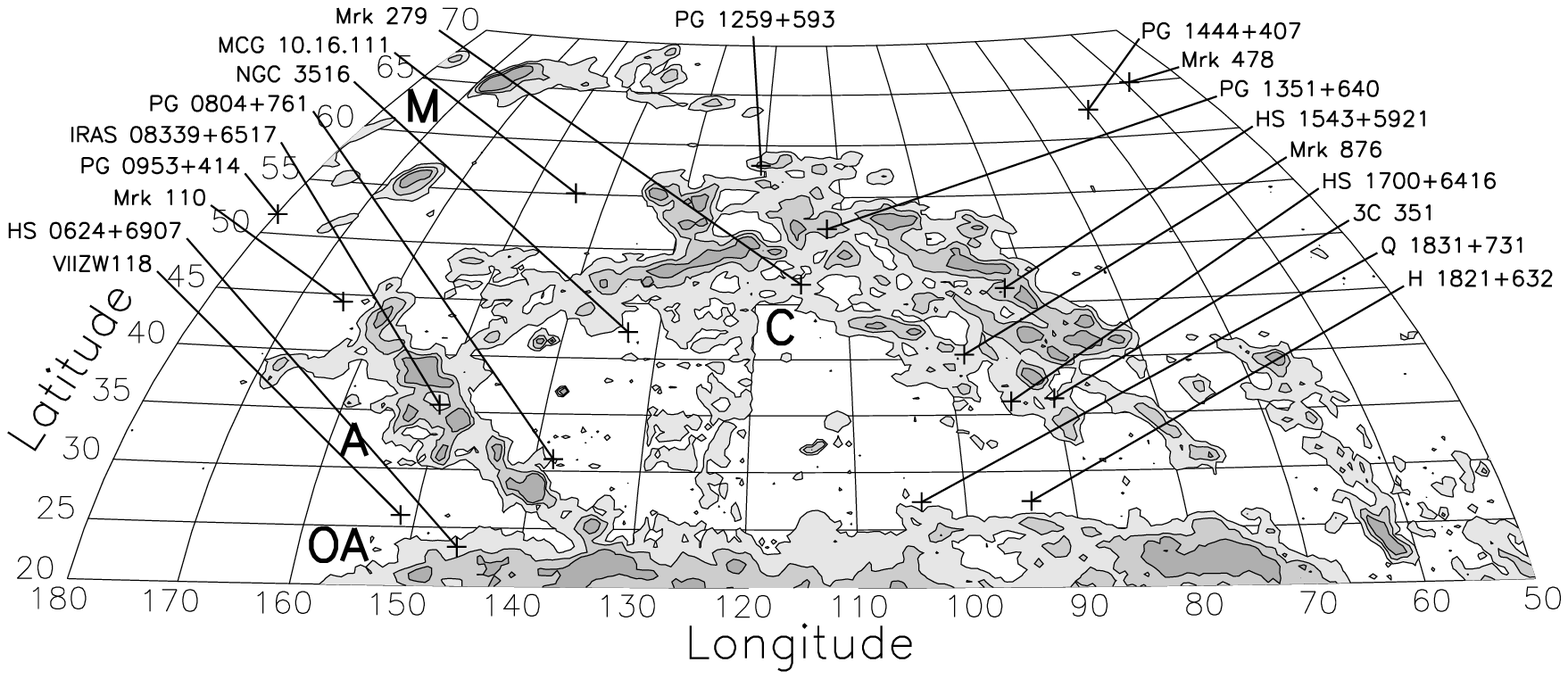}
\caption{Contours of \ion{H}{1}\ column density from the LDS (Hartmann 
\& Burton 1997) for high-velocity gas ($-210$ \kms $\leq 
V_{\rm LSR}\leq-90$ \kms) in the region $180\arcdeg\geq l \geq 50\arcdeg$ and
$70\arcdeg\geq b \geq 20\arcdeg$.  Contour levels are $N_{\rm HI} =1$, 3, and 
$6\times10^{19}$ cm$^{-2}$.  The locations of the 19 sight lines
in this region are labelled.  High-velocity \ion{Si}{3}\ absorption
is detected within the plotted velocity range for all 19 sight lines. 
}
\end{figure}

\begin{figure}
\figurenum{4}
\plotone{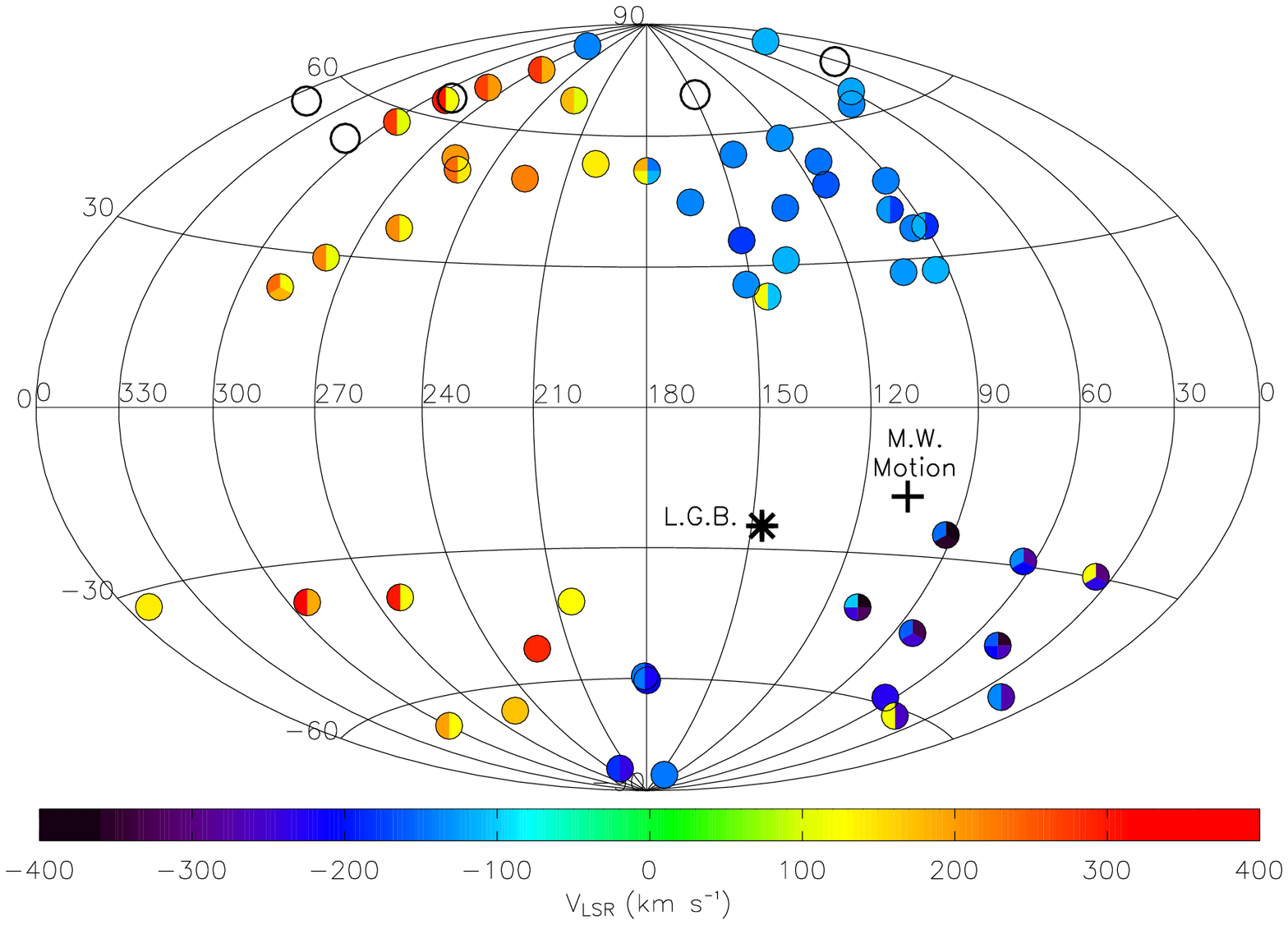}
\caption{All-sky Hammer-Aitoff projection in Galactic coordinates of the
locations of the 58 QSO sight lines analyzed for this study.  
The color scale indicates the velocity centroid of the \ion{Si}{3}
line for the HVC component(s) in the sight line.  For sight lines with multiple
HVC components, the circle is split into several colors representing the 
velocity centroid of each component.  Empty circles indicate sight lines 
where no high-velocity \ion{Si}{3}\ is detected.
We mark the Local Group barycenter ($\ell = 147\arcdeg$, $b = -25\arcdeg$;
Karachentsev \& Makarov 1996) by an asterisk, and note with a plus sign
the direction of the Milky Way motion ($V = 90$ km~s$^{-1}$, $\ell = 107\arcdeg$, 
$b = -18\arcdeg$; Einasto \& Lynden-Bell 1982) with respect to the Local Group
barycenter.}
\end{figure}

\begin{figure}
\figurenum{5}
\epsscale{0.85}
\plotone{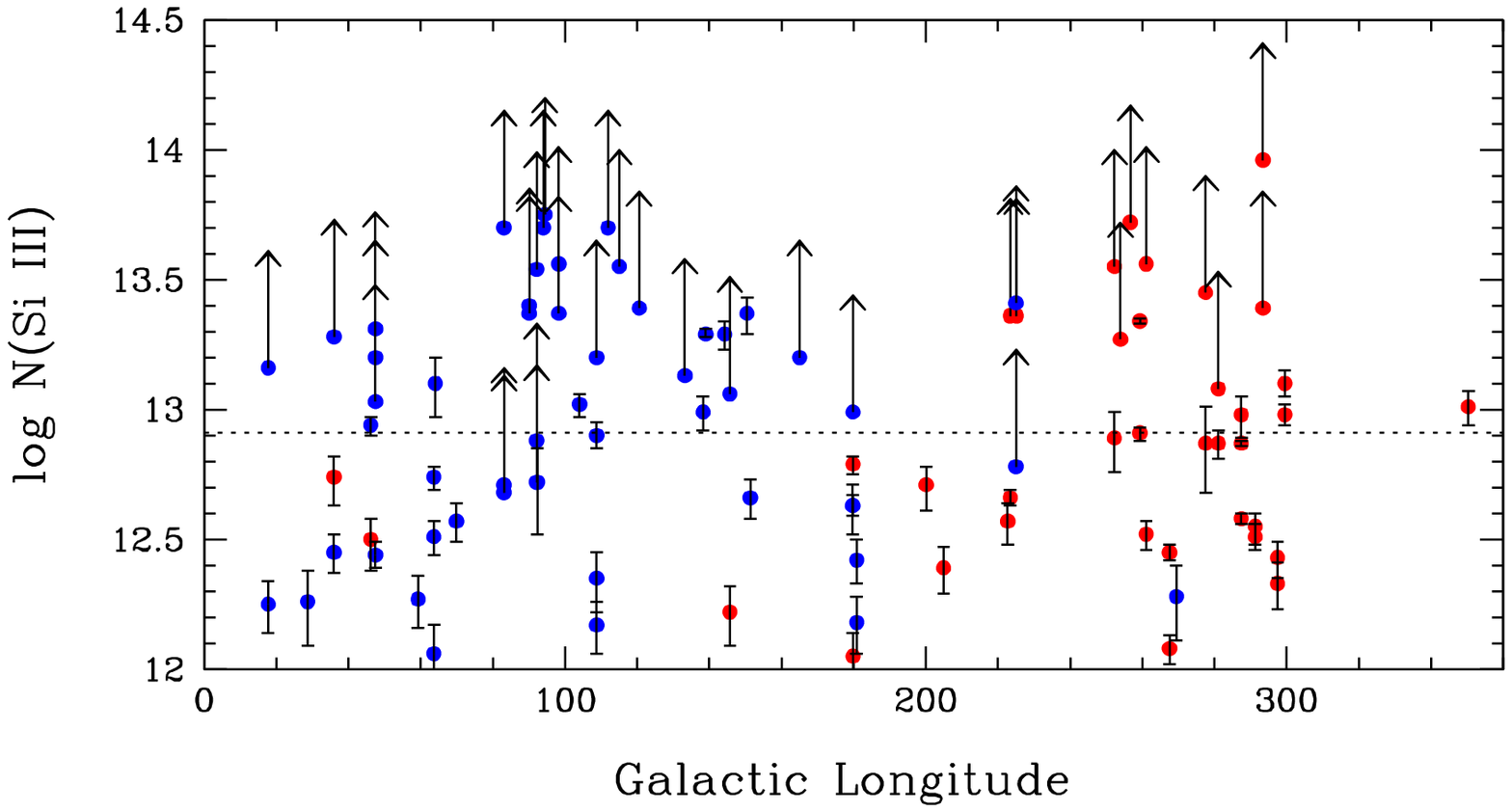}
\plotone{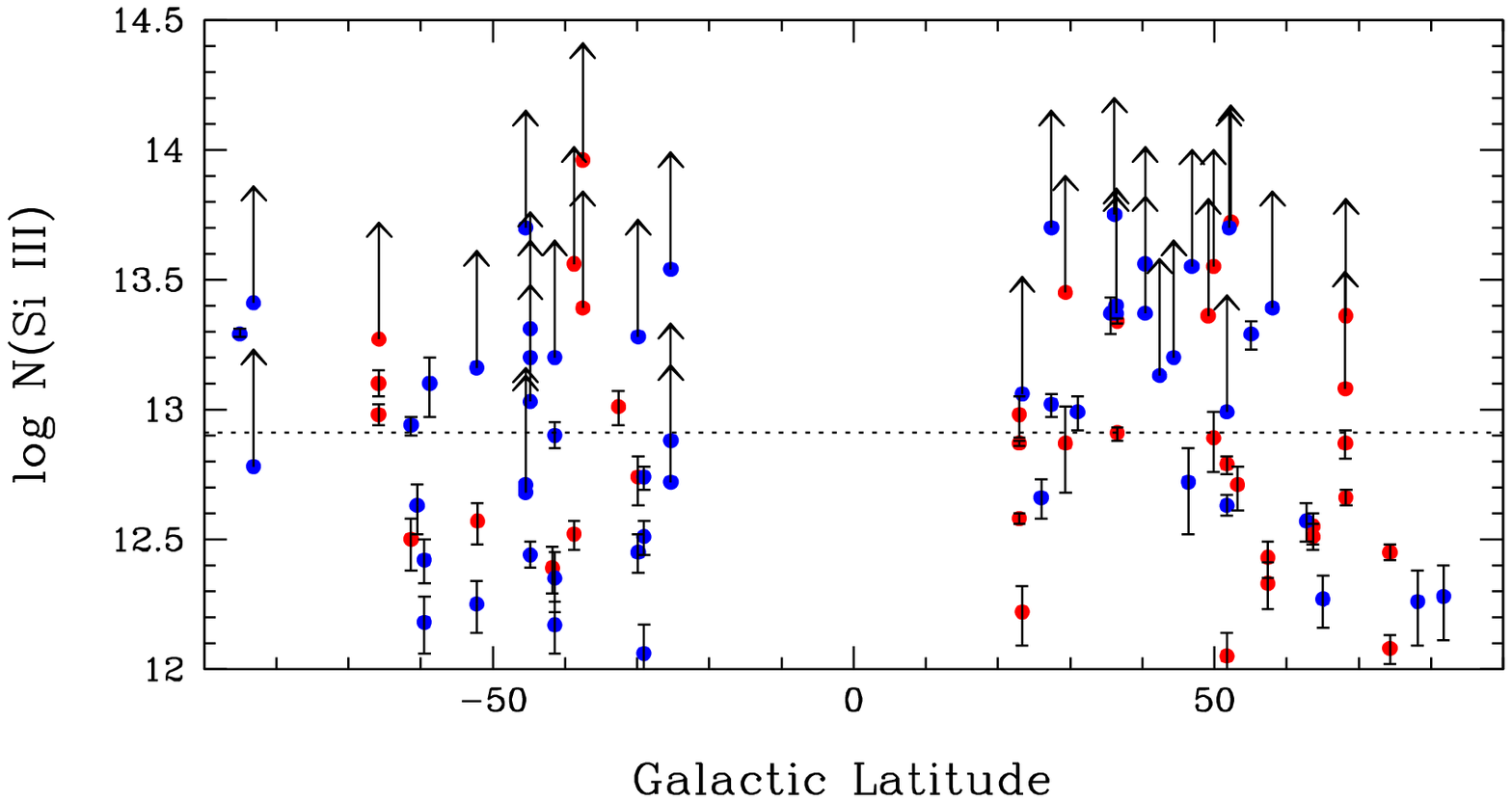}
\caption{Column densities N(cm$^{-2}$) of \SiIII\ vs.\ Galactic 
longitude and latitude.  Each point represents an individual velocity
component for the HVCs. The dashed line represents the median logarithmic
\SiIII\ column density of 12.91.  Lower limits reflect saturation in the 
strong \SiIII\ $\lambda 1206.5$ absorption line. Red-shifted and 
blue-shifted components are shown in appropriate colors.  No clear
trends are apparent, other than possible peaks in \SiIII\ 
(lower limits in column density) around $\ell = 80^{\circ}-120^{\circ}$ 
and $\ell = 260^{\circ}-300^{\circ}$.  
}
\end{figure}

\begin{figure}
\figurenum{6}
\epsscale{1.2}
\plotone{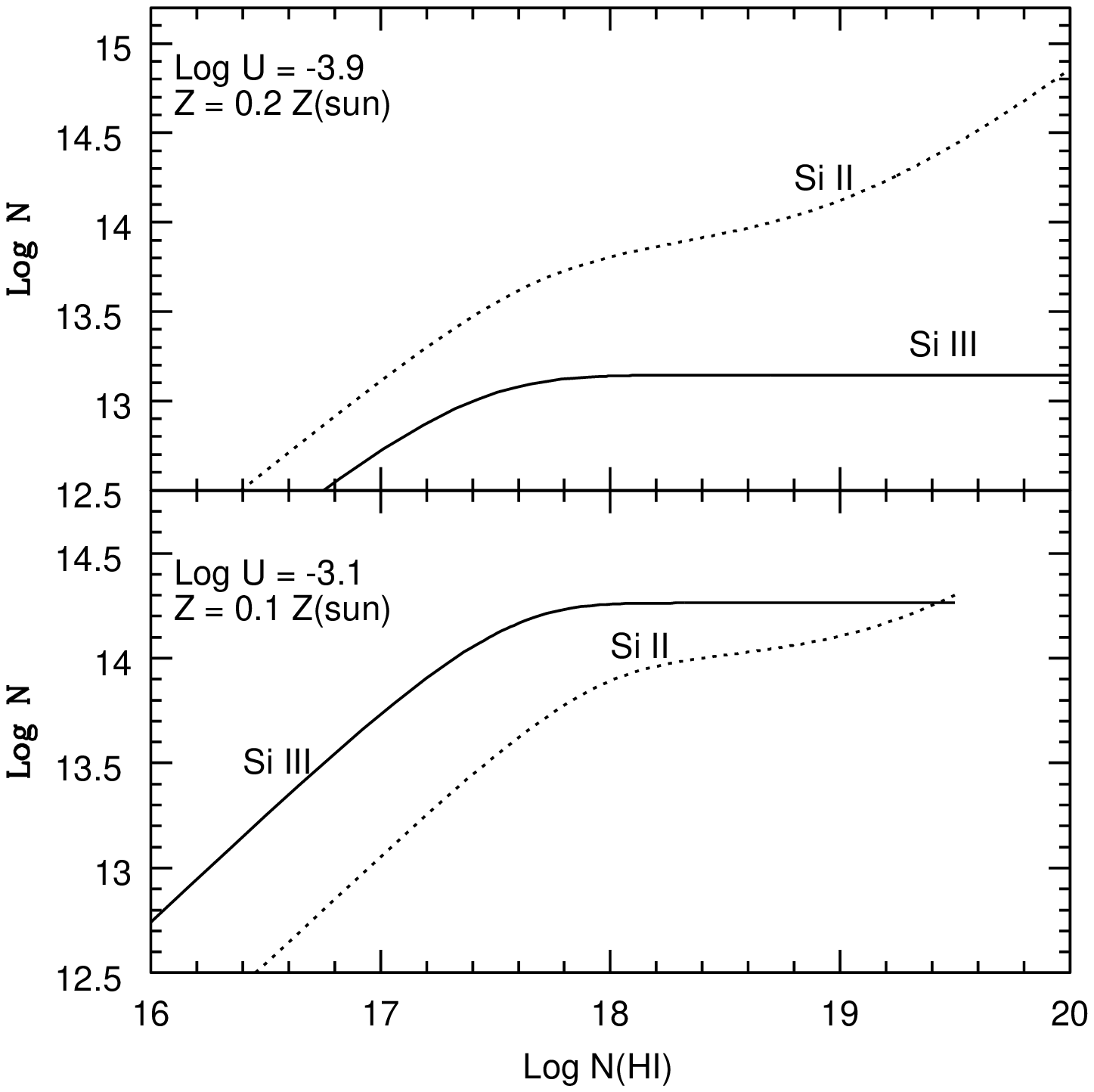}
\caption{Typical HVC photoionization models of \SiII\ and \SiIII\ column 
densities vs.\ depth into absorber as measured by N$_{\rm HI}$. Two panels 
show models with different metallicities and ionization parameters 
(a) $\log U = -3.9$, $n_H = 0.1$ cm$^{-3}$, and 20\% solar abundances; 
(b) $\log U = -3.1$, $n_H = 0.016$ cm$^{-3}$, and 10\% solar abundances.  
These models are similar to those in Shull \etal\ (2009) 
and suggest that some ionized HVCs with N$_{\rm SiIII} \geq10^{13.5}$ \cd\ 
could be detectable in 21-cm emission if log~\NHI $\geq 17.5$.  
}
\end{figure}

\end{document}